\documentclass[conference]{IEEEtran}

\usepackage{booktabs}
\usepackage{epstopdf}
\usepackage{graphicx}
\usepackage{amsmath}
\usepackage{subcaption}
\usepackage{comment}
\usepackage{amsmath}
\usepackage[english]{babel}
\usepackage[linesnumbered,algoruled,boxed,lined]{algorithm2e}
\usepackage{amsfonts}
\usepackage{amssymb}
\usepackage{amsthm}
\usepackage[colorinlistoftodos,prependcaption,textsize=tiny]{todonotes}

\begin{document}
\title{GreenDataFlow: Minimizing the Energy Footprint of Global Data Movement}

\author{\IEEEauthorblockN{MD S Q Zulkar Nine\textsuperscript{1}, Luigi Di Tacchio\textsuperscript{1}, Asif Imran\textsuperscript{1}, Tevfik Kosar\textsuperscript{1}, M. Fatih Bulut\textsuperscript{2}, Jinho Hwang\textsuperscript{2}}
\IEEEauthorblockA{\textsuperscript{1}Department of Computer Science and Engineering,
University at Buffalo,
Buffalo, New York\\
\textsuperscript{2}IBM TJ Watson Research Center, Yorktown Heights, New York\\
Email: \{mdsqzulk, luigidit, asifimra, tkosar\}@buffalo.edu, \{mfbulut, jinho\}@us.ibm.com}
}

\maketitle
\begin{abstract}
The global data movement over Internet has an estimated energy footprint of 100 terawatt hours per year, costing the world economy billions of dollars. The networking infrastructure together with source and destination nodes involved in the data transfer contribute to overall energy consumption. Although considerable amount of research has rendered power management techniques for the networking infrastructure, there has not been much prior work focusing on energy-aware data transfer solutions for minimizing the power consumed at the end-systems. In this paper, we introduce a novel application-layer solution based on historical analysis and real-time tuning called GreenDataFlow, which aims to achieve high data transfer throughput while keeping the energy consumption at the minimal levels. GreenDataFlow supports service level agreements (SLAs) which give the service providers and the consumers the ability to fine tune their goals and priorities in this optimization process. Our experimental results show that GreenDataFlow outperforms the closest competing state-of-the art solution in this area 50\% for energy saving and $2.5\times$ for the achieved end-to-end performance.
\end{abstract}

\section{Introduction}
\label{sec:introduction}
The era of artificial intelligence (AI) has made data the most important resource, in turn the efficient data handling is the key to use compute, network, and storage resources more effectively. Not like compute and storage resources, the network resource needs more sophisticated control as it involves the end-to-end efficiency. The annual data transfer rate over global IP networks has already exceeded zettabyte scale~\cite{Cisco_2016}. The energy footprint of this global data movement is estimated at more than 100 terawatt hours per year at the current rate, costing more than 20 billion US dollars annually to the world economy in addition to the environmental side effects~\cite{Gupta_2003, MINTS_2012, Mahadevan_2009, Cisco_2016, gao2012s}. This fact has resulted in considerable amount of work focusing on power management and energy efficiency in hardware and 
software systems~\cite{Brooks:2000:WFA:339647.339657, rawson2004mempower, zedlewski2003modeling, gurumurthi2002using, economou2006full,  fan2007power, rivoire2008comparison, koller2010wattapp, hasebe2010power, vrbsky2013decreasing, qureshi2009cutting} as well as on power-aware networking~\cite{Katz_2008, Mahadevan_2009, Greenberg_2009, Heller_2010, Goma_2011, fu2012frequency}. 

Majority of the existing work on power-aware networking focuses on reducing the power consumption on networking devices (i.e., routers, switches, and hubs). Gupta et al.~\cite{Gupta_2003} were amongst the earliest researchers to advocate conserving energy in the networking infrastructure. They suggested different techniques such as putting idle sub-components (i.e., line cards, etc.) to sleep~\cite{Gupta_2007}, which were later extended by other researchers. Nedevshi et al. proposed adapting the rate at which switches forward packets depending on the traffic~\cite{Nedevschi_2008}. 
IEEE Energy Efficient Ethernet Task Force proposed the 802.3az standards ~\cite{IEEE_802} for making Ethernet cards more energy efficient. They defined a new power state called Low-Power Idle (LPI) that puts the Ethernet card to low power mode when there is no network traffic.
Other related research in power-aware networking has focused on architectures with programmable switches~\cite{Greenberg_2008}, switching layers that can incorporate different policies~\cite{Joseph_2008}, and power-aware network protocols for energy efficiency in network routing~\cite{Barford_2008}.

The existing approaches suffer from the following drawbacks: (1) the solution is too costly (i.e., replacing all switches with energy efficient ones); (2) the solution is unpractical in the short term (i.e., replacing TCP with a more energy-efficient version); (3) the solution penalizes performance while increasing energy efficiency (i.e., sleeping some components while not in use). In this paper, we propose an application-layer solution called GreenDataFlow which is low cost, very easy and practical to deploy, and does not penalize the performance while increasing energy efficiency. With the added benefits and simplicity to adopt, service providers can directly benefit from GreenDataFlow as they can offer it as-a-service offering in their cloud platforms, while making sure that the SLA requirements of customers are satisfied using GreenDataFlow's SLA-based algorithms.

\begin{figure*}[t]
    \begin{centering}
    \begin{subfigure}[t]{0.40\textwidth}
        \includegraphics[keepaspectratio=true,width=70mm]{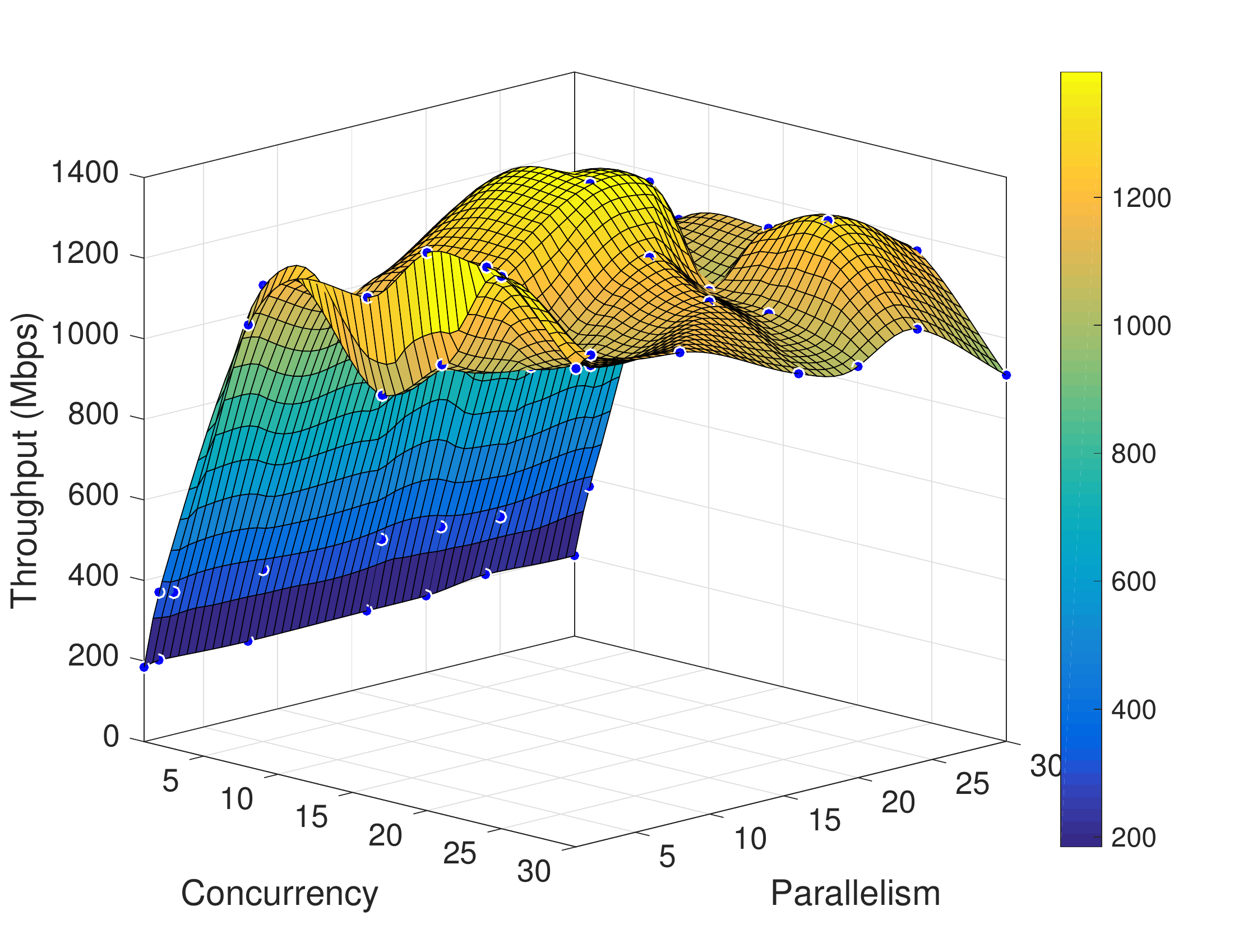}
        \caption{Achieved Throughput for different $cc$ and $p$}
    \end{subfigure}
    ~
    \begin{subfigure}[t]{0.40\textwidth}
        \includegraphics[keepaspectratio=true,width=70mm]{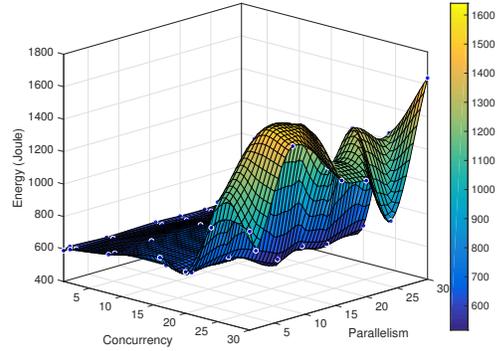}
        \caption{Energy Consumption for different $cc$ and $p$}
    \end{subfigure}
     \caption{Achieved throughput and energy consumption of a single transfer under different parameter combination. Surface interpolation is perform using piece-wise cubic spline. }
     \label{fig:sample_transfer}
     \end{centering}
 \end{figure*}

GreenDataFlow provides novel two-phase dynamic optimization models to minimize energy and increase throughput at the same time. It is based on mathematical modeling with offline knowledge discovery and adaptive online decision making. During the offline analysis phase, we analyze historical transfer logs to perform knowledge discovery about the characteristics of the past transfers with similar requirements. During the online phase, we use the discovered knowledge from the offline analysis along with real-time investigation of the network condition to optimize the protocol parameters for both minimal energy consumption and maximum transfer throughput. Our models use historical knowledge about the network and data to reduce the real-time investigation overhead while ensuring near optimal results for each transfer. Specifically our contributions in this paper are as follows:

 \begin{enumerate}
 
 \item GreenDataFlow achieves minimizing the energy footprint of an end-to-end big data transfer by operating in the application-layer, without any need to change the existing infrastructure nor the low-level networking stack, which makes integration of GreenDataFlow to existing applications easier.
 
\item GreenDataFlow integrates knowledge-based offline analysis with real-time tuning to achieve close-to-optimal end-to-end data transfer throughput while reducing energy consumption.
 

\item GreenDataFlow applies adaptive tuning in real-time and uses pre-computed mathematical optimization based decisions to provide faster convergence toward maximally achievable throughput. 


\item GreenDataFlow outperforms state-of-the-art solutions in this area in terms of accuracy, convergence speed, energy efficiency, and achieved throughput. Our experimental results show that GreenDataFlow outperforms the closest competing state-of-the art solution in this area up to 50\% for energy saving and up to $2.5\times$ for the achieved end-to-end performance. When we compare it with baseline cases (without any optimization), the energy savings go up to 80\% and performance improvement reaches $10\times$.

\end{enumerate}

The rest of the paper is organized as follows: Section \ref{sec:Problem Formulation} gives a formal definition of the problem; Section \ref{sec:challenges} discusses the challenges in optimization; Section \ref{sec:Proposed Model} presents our novel two-phase dynamic optimization model; Section \ref{sec:experiments} evaluates our model; Section \ref{sec:Related Work} describes the related work in this field; and Section \ref{sec:conclusion} concludes the paper.

\section{Problem Formulation}
\label{sec:Problem Formulation}
Large-scale data transfers can get suboptimal performance and high energy footprint in a long RTT WAN network due to the protocol inefficiency introduced in different layers.   
Changing the protocol stack requires low-level updates (e.g., modifications to TCP), and its adaptation by large-scale needs considerable time and effort.
Therefore, application level solutions are more lucrative and easy to deploy in the user space which makes the adaptation of the solution hassle-free.  
Application level data transfer protocol parameters (i.e., concurrency, parallelism, pipelining, and buffer-size) can have different impacts on transfer throughput and energy consumption of files with different sizes under certain network conditions. These parameters can be tuned to increase the data transfer throughput and decrease the energy footprint significantly. Figure~\ref{fig:sample_transfer} shows the achieved throughput and energy consumption of a single transfer under different parameter combinations. A short description of these parameters is given below.
 
\textbf{Concurrency} ($cc$) refers to the task level parallelism. It controls the number of server processes where each process can transfer an individual file. It can accelerate the transfer throughput when a large number of files need to be transferred. Concurrency can also take advantage of parallel file systems (e.g., GPFS, Lustre) with multiple concurrent servers and metadata management. 
 
Each server process can transfer a different portion of a file in parallel. We define the number of parallel streams for each server as \textbf{Parallelism} ($p$). It is a good choice to transfer medium and large files. We can get the full performance of parallelism with parallel file systems where files are divided and distributed on different disks. The total number of parallel data streams can be expressed as $(cc\times p)$. Increasing number of parallel data streams can increase the achievable throughput, however, excessive use of streams may lead to packet loss and force TCP to initiate slow-start phase that may lead to severe throughput loss.    

Control channel idleness is a major bottleneck to transfer a large number of files. After each file transfer, the server process sends an acknowledgment to initiate the next file transfer. This acknowledgment can take at least one Round-Trip-Time (RTT) between each transfer. This one RTT delay may seem innocent, however, it may hurt the overall throughput of a dataset containing a large number of small files significantly in a long RTT network. Because small files take short time to transfer and then each file needs to wait for one RTT to get acknowledgment from previous transfer. Moreover, TCP will shrink window size to zero if it detects data channel idleness. These issues can be solved by queuing multiple file transfer requests without waiting for the acknowledgments. This technique can transfer large number files like a single large file. We define the size of the outstanding file transfer request queue as \textbf{Pipelining} ($pp$). 

Energy consumption is a major concern in data centers and end-systems that perform very large scale data transfers. Minimizing the energy consumption can reduce the data center operating cost-effectively while utilizing the network links efficiently. Energy consumption can be measured using specialized hardware meters. However, we need an energy consumption model to estimate the real-time consumption that can be used to facilitate fine-grained power tuning. 
Energy consumption can be estimated from the current load of the system, such as - CPU utilization ($\mu_{cpu}$), memory utilization ($\mu_{mem}$), disk usage ($\mu_{disk}$), and network interface card utilization ($\mu_{nic}$). In literature, there exist many models to predict the actual energy consumption using these load information. We have used a linear model to predict power consumption, which is presented in Section~\ref{sec:offline_optimization}.

Our aim is to perform an energy constrained optimization of the data transfer performance. End users or data center operations team may set these constrained optimization problem based on their requirements and priorities. In this work, we introduce easy to describe energy-aware Service Layer Agreement (SLA) categories that a user or an administrator can initiate. SLA is a contract between the user and the service provider where these two parties agree on service quality and specific rights of both parties. It may include the description of services agreed to be provided, monitoring and reporting of quality of service (QoS) matrices, the consequence of not meeting the requirements, and escape clause (the situation where service guarantee promised does not apply). For energy efficient transfers, SLA can be defined in three major ways: (1) Throughput guarantee \textbf{(Type-T)}, (2) Constraint over total energy usage \textbf{(Type-E)}, and (3) Constraint over instantaneous power consumption \textbf{(Type-P)}. We translate the SLAs into appropriate optimization problems and solve them off-line. These SLA categories are explained below. 

\textbf{\textit{(1) Throughput Guarantee: }}A user may need an overall throughput guarantee that means the achievable throughput $T_{act}$ must be at least the throughput specified in SLA, $T_{sla}$. In such case, the service provider would try to maintain the SLA requirement using as less energy, $E$ as possible. Therefore, the optimization problem can be expressed as :
\begin{equation}
\begin{array}{ll}
\underset{ \{cc,p,pp\} }{\mathrm{argmin}} &   (E) \\
\text{subject to.} & T_{act} \geq T_{sla}
\end{array}
\end{equation}

\textbf{\textit{(2) Constraint over Total Energy Usage: }} User or administrator may want to minimize energy cost by putting a constraint over the total energy consumption and request for the best possible throughput under this constraint. Therefore, actual total energy consumption ($E_{act}$) can be constrained by energy consumption level specified in SLA, ($E_{sla}$). The optimization problem is to maximize throughput, $T$ under a specified energy constraint and can be expressed as:

\begin{equation}
\begin{array}{ll}
\underset{ \{cc,p,pp\} }{\mathrm{argmax}} & \frac{1}{\tau_f - \tau_s} \displaystyle\int_{\tau_s}^{\tau_f} T \\
\text{subject to.} & E_{act} \leq E_{sla}.
\end{array}
\end{equation}
Where $\tau_s$ and $\tau_f$ are the starting time and the finish time respectively. 

\textbf{\textit{(3) Constraint over Instantaneous Power Consumption: }} Sometimes spikes in instantaneous power consumption can be very expensive, as power grid imposes a high penalty for such spikes. The user may want to put constraint over instant power consumption. That means the user wants to maximize the throughput with a guarantee that instant power consumption should not exceed the power limit mentioned in the SLA. To do that we need to model the instant power, $\varphi$ first based on resource utilization. 

\begin{equation}
\varphi = f(\mu_{cpu},\mu_{mem},\mu_{disk},\mu_{nic})
\end{equation}

\begin{equation}
\begin{array}{ll}
\underset{ \{cc,p,pp\} }{\mathrm{argmax}} & \frac{1}{\tau_f - \tau_s}  \displaystyle\int_{\tau_s}^{\tau_f} T \\
\text{subject to.} & \varphi_{i} \leq \varphi_{sla}.
\end{array}
\end{equation}

For all three cases, we need to schedule the compute resources to the server processes in a way that minimizes the energy consumption or keep it below the SLA constraint and simultaneously increase throughput. Dynamic load balancing among the parallel streams also helps to alleviate extra load from the congested streams.   

Several assumptions are made to design our optimization model. Those are explained below:

\theoremstyle{remark}
\newtheorem{assumption}{Assumption}
\begin{assumption}\label{as:1}
For disk-to-disk transfer, the achievable throughput ($T_{act}$) should be bounded by the bottleneck resource that can be end-to-end link bandwidth or disk read/write speed ($v_{read}$ and $v_{write}$) at the source and destination. 

\begin{equation}
T_{act} \leqslant \textrm{min} \{BW,v_{read},v_{write} \}
\end{equation}

\end{assumption}

\begin{assumption}\label{as:2}
We aim to optimize the application-level parameters and our solution is agnostic of the underlying file system. On the other hand, if the target data is stored on a parallel file system, that would benefit more from our optimization. 
%
\end{assumption}

\begin{assumption}\label{as:3}
Our model is also agnostic of the underlying reliable transport protocol. Any reliable transport protocol can work with our model. In this paper, we apply our optimizations to GridFTP~\cite{Liu:dataTxFr} which is based on TCP. GridFTP is widely used in the scientific community, and it supports easy tuning of parameters such as parallelism, concurrency, and pipelining. 
\end{assumption} 

\begin{assumption}\label{as:4}
Our model can work with/without Remote Direct Memory Access (RDMA) technology. Surely RDMA technology can reduce the overhead introduced by the kernel copy. However, it requires special hardware (i.e., RDMA NIC) support which can be a deployment barrier for the RDMA technology in end users. 
\end{assumption}

\section{Challenges}
\label{sec:challenges}
The main challenge is the dynamic nature of the shared network links.
Achievable transfer throughput can change from time to time during large-scale data transfers. To get required performance, the transfer needs online update and micro-tuning of the parameters, and real-time resource scheduling. 
Searching for updated parameters (i.e., solving the optimization problems mentioned in Section~\ref{sec:Problem Formulation}) during the transfer is expensive and introduces delay or transfer may continue with suboptimal parameters until search finishes. 
Both can introduce throughput loss to the transfer. 
To address the issue we use offline analysis of historical transfer logs to solve optimization problems. In this case, we need to solve the optimization problem for all possible SLA requirements which may be infeasible and take a considerable amount of memory to store solutions. We introduce an intelligent way to overcome the issue (discussed in Section~\ref{subsec:distributed_approach}). We store the solution as the key-value pair. During the actual transfer, we periodically check network conditions and if needed we can look up for the new parameter settings in constant time. 
We introduce a two-phase optimization model which combines the benefits of offline analysis and online tuning to ensure the SLA requirements. 
%
We introduce some design-specific challenges below.

\theoremstyle{remark}
\newtheorem{challenge}{Challenge}
\begin{challenge}\label{ch:1}
\textbf{(Cost of offline analysis)} Offline analysis itself introduces extra energy and latency cost. If not performed efficiently, it may defeat the purpose of offline analysis, when the combined cost of offline analysis $Cost$(\textit{Offline}) and the cost of actual data transfer, $Cost(Tr_{tuned})$ exceed the cost of the transfer without optimization, $Cost(Tr_{no\_opt})$. So the constraint can be imposed as:

\begin{equation}
Cost(\textit{Offline}) + Cost(Tr_{tuned}) < Cost(Tr_{no\_opt}) 
\end{equation}
\end{challenge}

\begin{challenge}\label{ch:3}
\textbf{(Packet loss)} Any packet loss can reduce the TCP window size (cwnd) significantly (depending on the TCP variant) which reduces the achievable transfer throughput. Packet loss can happen for a range of reasons, such as: (1) congestion, (2) bottleneck network devices with low capacity (routers/switch/middleboxes), (3) software bugs in network devices, and (4) faulty links. We need a mechanism to detect the static packet losses (2-4) and eliminate them manually. We also need a mechanism that can efficiently signal congestion beforehand to avoid the TCP window size reduction.  
\end{challenge}

\begin{challenge}\label{ch:4}
\textbf{(Fairness)} We need to ensure bandwidth usage fairness among the contending transfers. This means our model should not be too aggressive to increase the achievable throughput for only a specific set of users or for specific transfers. 
\end{challenge}

\begin{challenge}\label{ch:6}
\textbf{(Resource scheduling to sender/receiver processes)}  
We have explained in Section~\ref{sec:Problem Formulation} that the energy consumption can be modeled as a function of resource utilization. Here, the goal is to find the optimal number of server processes ($cc$) with multi-threaded sockets ($p$) and pipelining ($pp$) and then schedule the compute resources to the concurrent processes in a way to maintain the SLA agreement. Current CPU schedulers are not designed to meet such optimization goal. As an example, Linux uses Completely Fair Scheduling (CFS) and Real-time scheduler (RT) to schedule the CPU. CFS tries to mimic the perfectly fair scheduling. 
CFS achieves fairness by calculating the time slice as a fraction of the total number of running processes. 
In our case, not all the concurrent processes are pumping data at the same rate and some server processes may have more workload than the others. We may need to limit the CPU usage of certain server processes to maintain a steady power consumption rate. Current CFS scheduler cannot make such forced external modification. 
Cgroup is a Linux kernel tool that can provide more fine-grained external control over the resource scheduling. 
\end{challenge}

\section{Model Design}
\label{sec:Proposed Model}
\setlength\belowcaptionskip{5ex}
\begin{table}[t]
\small
	\begin{centering}
		\begin{tabular}{ c@{\hskip 0.5cm}l}
			\hline
			{\bf Symbol} &  {\bf Description}   \\
			\hline

			$T$	&  Throughput of the data transfer 	 	\\
			$E$	&  Energy consumption of the transfer 	 	\\	
            $\varphi$	&  Instantaneous power consumption \\
			$cc$	&  Concurrency 	 	\\
			$p$	&  Parallelism	 		\\
			$pp$	&  Pipelining 		\\
            $bs$   & End system buffer size \\
            $\mathbb{S}$ & Total number of streams, $\mathbb{S} = cc \times p$  \\ 
            $\theta$	&   Set of parameters, $\theta = \{cc,p,pp,bs\}$  \\
            $BW$   &   Bandwidth  \\
            $\eta$   &   Energy efficiency, $Data/ Energy$ \\
            $cc_{limit}$, $p_{limit}$    &   User limit on corresponding parameters \\
            $\tau$ & Time \\
            $\epsilon$   &   Throughput fluctuation tolerance bound \\
            $\mu$   &   Utilization of corresponding resource \\
            $\mu_{all}$  & Set of all resource utilization \\
            $I$   & Interpolant of Throughput or Energy log \\
 			
            \hline
            
		\end{tabular}
		\caption{Description of different symbols} 
        \label{tab:symbol}
	\end{centering}
\end{table}

We introduced two models to address the problem: (1) distributed and (2) centralized. In the distributed solution, there is no centralized control and all the users run their own optimization model and converge to receive a fair share of network throughput. However, in this situation, many users may end up solving the same or similar optimization problems. On the other hand, the centralized solution takes the burden of running optimization from the end users and performs optimization based on historical logs and shares it with the end users. The centralized scheduling is efficient for intra/inter data center large-scale transfers due to the fact that it can be integrated with Traffic Engineering (TE) module for joint optimization of throughput and energy consumption.  
The centralized scheduler can optimize data center energy consumption more efficiently than the distributed solution. 

\subsection{Distributed Approach}
\label{subsec:distributed_approach}
\begin{figure}[t]
\begin{centering}
\includegraphics[width=86mm, height=40mm]{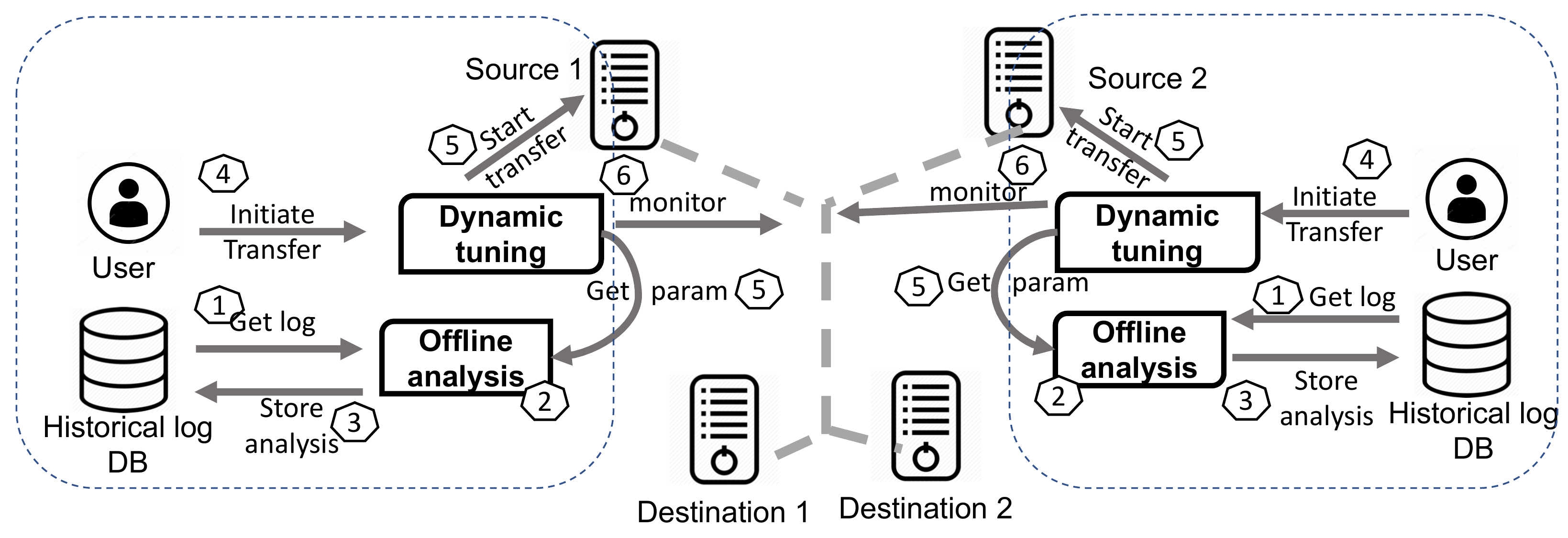}
	\end{centering}
\caption{Distributed approach overview.} 
\label{fig:distribute_overview}
\end{figure}

Our distributed approach has two phases: (1) offline optimization, (2) dynamic tuning. Offline optimization takes historical data transfer logs as input and solves the SLA based optimization problem beforehand. We choose this strategy so that optimization does not introduce latency during the actual transfer. The cost of offline optimization can be amortized over many subsequent transfers. As we have explained in Section~\ref{sec:Problem Formulation} different SLA requirements have different optimization objective. However, all of them are actually looking for optimal application-level parameters. In a shared environment, where dynamic fluctuation of traffic is very common, the static parameters from the offline analysis may become sub-optimal during the transfer. A real-time tuning is necessary to cope with such fluctuation as we have to maintain strict SLA constraint. An overview of the model is depicted in Figure~\ref{fig:distribute_overview}.

\subsubsection{Offline Optimization}
\label{sec:offline_optimization}
The Offline optimization can be broken down into parts: (1) storing historical logs, (2) external load modeling, (3) clustering and data interpolation, (5) constrained optimization, and (6) energy modeling. 

\vspace{1mm}
\noindent {\textbf{Step 1 -- Storing Historical Logs:}} 
Historical data transfer logs are collected periodically during the transfer and stored in a log server.
These logs collect information about the  network characteristics (e.g., round trip time, buffer size, queuing delay, packet loss rate), application level parameters ($cc$, $p$, $pp$), end system resource information (e.g., CPU, memory, NIC), data set information (e.g., size, number of files/objects), energy-related information (e.g., CPU utilization, memory utilization, NIC card utilization, disk I/O utilization).
These logs provide insight to optimize transfers energy-efficiently under different circumstances. 
Still, logs are prone to errors (e.g., recorded achieved throughput may be greater than the actual link bandwidth) and may have missing values. Standard data interpolation techniques are used to predict those values. A preprocessing module takes care of such case.

\vspace{1mm}
\noindent {\textbf{Step 2 -- External Load Modeling:}} 
In a shared network environment, data transfer task has to compete with other contending transfers. Contending transfers can be a mixture of known incoming/outgoing transfers in both source/destination and completely unknown transfers. Some bandwidth may be wasted by TCP congestion and slow start. However, the optimal number of streams can offset slow start and congestion loss significantly. We can model the achievable throughput of a data transfer as:

\begin{equation}
\begin{aligned}
T_{act} ={} & BW - \sum T_{ext\_known} - \sum T_{ext\_unknown}\\
 	        &  - \sum \delta_{slow\_start} - \sum \delta_{congestion} 
\end{aligned}
\label{eq:tact}
\end{equation}

As we have periodically collected logs, it is easy to estimate the $\delta_{slow\_start}$ and $\delta_{congestion}$ just analyzing the achieved throughput of the subsequent time intervals. We can also estimate the combined throughput of known transfers. Therefore, from Equation~(\ref{eq:tact}), we can get a rough estimate of the combined throughput of the unknown external traffic.  

\vspace{1mm}
\noindent {\textbf{Step 3 -- Clustering and Interpolation:}} 
Similar types of transfers can be optimized using similar parameter combinations.
Categorizing logs into groups based on their similarity could provide us a more structured view of the log information. After analyzing the logs we come to the conclusion that some parameters have direct precedence over other parameters. We use {\em Hierarchical Agglomerative Clustering}~\cite{murtagh2014ward} which is the most suitable clustering technique for such cases. 

We are interested to find the optimal parameters under different external traffic load. In Step-2 we get a rough estimate of external traffic. Our experiment shows that achievable throughput, $T_{act}$ and energy consumption, $E_{act}$ can be directly impacted by the application level parameters, $\theta[:]$ = \{$cc$, $p$, $pp$\}. The relation is strictly non-linear and follows a continuous cubic pattern. Therefore, we modeled both throughput and energy using piece-wise cubic spline interpolation (Figure~\ref{fig:sample_transfer}). This technique stitches multiple cubic functions with smoothness guarantee up to the second derivative. All the continuity constraints and the smoothness constraints are linear. Therefore, the coefficients can be computed by solving the system of linear equations. Interpolants can be written for each cluster $c_{i}$ as:

\begin{equation}
\begin{split}
T_{c_{i}} = I_1(p, cc, pp, bs) \\
E_{c_{i}} = I_2(p, cc, pp, bs)
\end{split}
\end{equation}

\vspace{1mm}
\noindent {\textbf{Step 4 -- Constrained Optimization:}} 
An overview of constrained optimization for different types of SLA is explained in Section~\ref{sec:Problem Formulation}. Running those optimizations in real-time may add extra latency during the transfer. One may argue that the real-time optimization can be performed concurrently with the transfer. Still, the transfer would run under sub-optimal solution until optimization finishes. Pre-computing these optimizations during offline phase can have two major benefits: (1) it eliminates any real-time latency for optimal parameters, and (2) these precomputed results can be reused for many subsequent transfers, which can effectively amortize the initial cost of analysis. As the user can put constraints over throughput or energy or instant power consumption, during offline phase we would have to solve the optimization problem for all possible SLA values of throughput or energy or power, which is not feasible. We solve this issue by intelligently analyzing the historical logs so that we can eliminate infeasible SLA constraint values. For example, to perform a transfer we need a minimum level of energy consumption, therefore, any SLA value below the threshold is infeasible. We can set a feasible region of SLA values. We also observed that, when two SLA values are close, they produce similar solutions. Therefore, instead of solving the optimization problem for all possible SLA values we partition the SLA feasible region based on historical log data and solve one optimization problem for each partition.    
This approach can give us a discrete number of SLA levels for both throughput ($k_{th\_level}$), energy ($k_{e\_level}$), and power consumption ($k_{p\_level}$). 

\vspace{1mm}
\noindent {\textbf{Step 5 -- Energy model:}} 

\begin{figure}[t]
\begin{centering}
\includegraphics[keepaspectratio=true,angle=0,width=70mm]{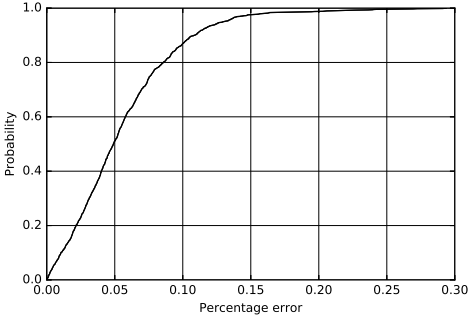}
	\end{centering}
\caption{Regression error CDF of power model.} 
%
%
\label{fig:regression_cdf_error}
\end{figure} 
 
When multiple server processes are involved, the total energy consumption of the end system can be estimated as the summation of energy consumption of the concurrent server processes, $e_{i}$ (energy consumption of $i$-th process). 

\begin{equation}
E_{\tau} =\sum_{i=1}^{cc} e_{i}(\mu_{cpu,i,\tau},\mu_{mem,i,\tau},\mu_{disk,i,\tau},\mu_{nic,i,\tau})
\end{equation}

In order to estimate the energy consumption of the end systems during the data transfers, we built a linear regression model with the following input features: CPU utilization, memory usage, number of disk reads and writes, number of bytes read and written to disk, number of bytes sent and received over the network, and number of packets sent and received. We collected 4467 samples by monitoring a workstation under different CPU and network loads, and measured the actual power consumption of the machine using a Yokogawa WT210 power meter. We used 70\% of the samples for training and 30\% for testing. Figure \ref{fig:regression_cdf_error} shows the CDF of the prediction error for the test set. As the figure shows, most predictions (>90\%) suffer from a very small error (<10\%).


\IncMargin{1em}
\begin{algorithm}[t]
\small
	\SetKw{in}{in}
	\SetKwData{Left}{left}
    \SetKwData{This}{this}
    \SetKwData{Up}{up}\SetKwData{Log}{Log}
    \SetKwData{ismedian}{$e_{s,median}$}\SetKwFunction{Median}{Median}
    \SetKwFunction{Union}{Union}
    \SetKwFunction{translateSLA}{translate\_SLA}
    \SetKwFunction{checkQueingDelay}{measure\_queuing\_delay}
    \SetKwFunction{estimateNewParam}{get\_params}
    \SetKwFunction{getPacketLossRate}{get\_packet\_loss\_rate}
    \SetKwFunction{measureResourceUtilization}{get\_resource\_utilization}
    \SetKwFunction{updateResourceGroups}{update\_resource\_groups}
    \SetKwFunction{checkStreamPerformance}{check\_stream\_perf}
    \SetKwFunction{redistributePipelining}{redistribute\_pipelining}
    \SetKwFunction{opportunisticIncrease}{opportunistic\_increase}
    \SetKwFunction{opportunisticDecrease}{opportunistic\_decrease}
    \SetKwFunction{reduce}{reduce}
    \SetKwFunction{increase}{increase}
    \SetKwFunction{decrease}{dec}
    \SetKwFunction{FindClosestSurface}{FindClosestSurface}
    \SetKwFunction{GetOptimalParam}{GetOptimalParam}
    \SetKwFunction{DataTransfer}{data\_transfer}
    \SetKwFunction{backoffcontrol}{back\_off\_control}
    \SetKwFunction{append}{append}
    \SetKwFunction{remove}{remove}
    \SetKwFunction{QueryDB}{QueryDB}
    \SetKwFunction{Sort}{Sort}
    \SetKwFunction{GetSamples}{GetSamples}
	\SetKwInOut{Input}{input}\SetKwInOut{Output}{output}
    \tcp{Expected Energy Efficiency, $\eta = Data_{total}/E_{sla}$; Queuing Delay, $Q_{rtt}$; number of streams, $\mathbb{S} = cc \times p$}
	\BlankLine
    $SLA\_type$ $\leftarrow$ \translateSLA{$SLA$} \\
    $\theta_{initial}$ $\leftarrow$ \estimateNewParam{$req$, $TYPE=median$} \\
    \updateResourceGroups{$E_{sla}$, $TYPE=median$} \\
    \DataTransfer{$req$, $\theta_{initial}$} \\
    \textbf{Periodically check:} \\
    \uIf{$SLA\_type$ $==$ \textsf{'Energy Constraint'}}{
    	\uIf{ $Data_{curr}/E_{curr}$ $\leq$ $\eta$}{ 
        	$E_{left} \leftarrow E_{sla} - E_{curr}$ \\
            \textbf{if} $T_{curr} \le T_{pred} - \epsilon:$ \backoffcontrol{} \\
              
            $\theta_{new}$ $\leftarrow$ \estimateNewParam{$Q_{rtt}$, $th_{prev}[i:j],$ $E_{lelf}$} \\  
        }
    	\Else {
        	\increase(cc\_limit,p\_limit, $\alpha_{cc}$, $\alpha_{p}$) \\
        	\opportunisticDecrease{$E_{left}$} \\
            $\theta_{new}$ $\leftarrow$ \estimateNewParam{$Q_{rtt}$, $th_{prev}[i:j],$ $E_{lelf}$} \\
        }
    }
    
    \uElseIf{$SLA\_type$ $==$ \textsf{'Throughput Guarantee'}}{
    	\If{ $T_{curr}$ $\neq$ $T_{sla} \pm \epsilon$}{ 
            $T_{goal} \leftarrow T_{sla} + (T_{sla} - T_{curr}) $ \\
        	\textbf{if} $T_{curr} \le T_{sla} - \epsilon:$ \backoffcontrol{} \\
            \textbf{else:} \increase(cc\_limit,p\_limit, $\alpha_{cc}$, $\alpha_{p}$) \\
            \opportunisticIncrease{$T_{goal}$} \\
            $\theta_{new}$ $\leftarrow$ \estimateNewParam{$Q_{rtt}$, $th_{prev}[i:j],$ $T_{goal}$} \\
            
        }
        
    }
    
    \uElseIf{$SLA\_type$ $==$ \textsf{'Power Constraint'}}{
    	\updateResourceGroups{$\varphi_{sla}$, $TYPE = fixed$} \\
    	\uIf{ $T_{curr}$ $\leq$ $T_{pred} - \epsilon$}{ 
        	$T_{goal} \leftarrow T_{pred} + (T_{pred} - T_{curr})$ \\
            \backoffcontrol{} \\
              
        }
        \Else{
               \increase($cc\_limit$, $p\_limit$, $T_{goal}$) \\
        } 
        $\theta_{new}$ $\leftarrow$ \estimateNewParam{$Q_{rtt}$, $th_{prev}[i:j],$ $T_{lelf}$} \\
    }
    $\mu[1:cc][:]$ $\leftarrow$ \measureResourceUtilization{} \\
    \updateResourceGroups{$E_{left}$} 
    \BlankLine
    
    \textbf{Periodically call:} \\
    $SSet\_low, th[1:\mathbb{S}]$ $\leftarrow$ \checkStreamPerformance{$cc,p$} \\ 
    \redistributePipelining{$SSet\_low$, $th[1:\mathbb{S}], cc, p$} \\
    \textbf{if} $required$ \textbf{:} \updateResourceGroups{$E_{left}$} 

\caption{Dynamic Tuning (Distributed)}
\label{algo:dynamic_tuning_distributd}
\end{algorithm}\DecMargin{1em}

\IncMargin{1em}
\begin{algorithm}[t]
\small
	\SetKw{in}{in}
	\SetKwData{Left}{left}
    \SetKwData{This}{this}
    \SetKwData{Up}{up}\SetKwData{Log}{Log}
    \SetKwData{ismedian}{$e_{s,median}$}\SetKwFunction{Median}{Median}
    \SetKwFunction{Union}{Union}
    \SetKwFunction{translateSLA}{translate\_SLA}
    \SetKwFunction{checkQueingDelay}{measure\_queuing\_delay}
    \SetKwFunction{estimateNewParam}{get\_params}
    \SetKwFunction{getPacketLossRate}{get\_packet\_loss\_rate}
    \SetKwFunction{measureResourceUtilization}{get\_resource\_utilization}
    \SetKwFunction{updateResourceGroups}{update\_resource\_groups}
    \SetKwFunction{checkStreamPerformance}{check\_stream\_perf}
    \SetKwFunction{redistributePipelining}{redistribute\_pipelining}
    \SetKwFunction{opportunisticIncrease}{opportunistic\_increase}
    \SetKwFunction{reduce}{reduce}
    \SetKwFunction{increase}{inc}
    \SetKwFunction{decrease}{dec}
    \SetKwFunction{FindClosestSurface}{FindClosestSurface}
    \SetKwFunction{GetOptimalParam}{GetOptimalParam}
    \SetKwFunction{DataTransfer}{data\_transfer}
    \SetKwFunction{backoffcontrol}{back\_off\_control}
    \SetKwFunction{append}{append}
    \SetKwFunction{remove}{remove}
    \SetKwFunction{QueryDB}{QueryDB}
    \SetKwFunction{Sort}{Sort}
    \SetKwFunction{GetSamples}{GetSamples}
	\SetKwInOut{Input}{input}\SetKwInOut{Output}{output}
    
    \SetKwProg{proc}{procedure}{}{}
    \proc{\backoffcontrol{} } {
    	$Q_{rtt}$ $\leftarrow$ \checkQueingDelay{} \\
        $pkt\_loss\_rate$ $\leftarrow$ \getPacketLossRate{} \\
            \If{ $Q_{rtt}$ $\ll$ $Q_{rtt}.expected$)} {
            	\reduce{$p\_limit, \beta_1$}  
            }
            \If{ $pkt\_loss\_rate$ $\ll$ $pkt\_loss\_rate.threshold$} {
            	\reduce{$cc\_limit, \beta_2$}  
            }
    }

\caption{Stream Back off Control Algorithm}
\label{algo:backoff_control}
\end{algorithm}\DecMargin{1em}

\subsubsection{Dynamic Tuning}
Dynamic tuning is the heart of the protocol. This is the real-time monitoring of the health of the data transfers, simultaneously it controls the aggressiveness of the protocol (fairness constraint), while ensuring strict SLA requirement. As we have three different categories of SLAs, we need three different strategies as well. However, the core control is mostly similar. An overview of the tuning module is introduced in Algorithm~\ref{algo:dynamic_tuning_distributd}.
To achieve energy efficiency we need to restrict the resource utilization of the transfer processes. We used cgroup to predefine some resource groups where resource utilization can be restricted up to defined levels. Transfer processes can be assigned to those resource group in real-time. 
Large-scale transfers take a long time, therefore, network load fluctuation is a reality. 
During the transfer, there may be some time-interval when the network becomes congested due to newly initiated external transfers and achievable throughput may drop up to a certain level. To maintain fairness, we introduce a back-off control to reduce parameters (Algorithm~\ref{algo:backoff_control}). When queuing delay drops for a certain amount of time, it reduces parallelism level by predefined $\beta_1$ (Line 4-6). When it detects a significant increase in packet loss rate, it reduces the concurrency level by predefined $\beta_2$ (Line 7-9). During these time intervals achieved throughput may go below the throughput mentioned in SLA, which must be compensated to guarantee the SLA requirement. 
%
We introduce an opportunistic strategy that is - whenever possible (time intervals with low external load) target for a solution better than SLA requirement so that we have enough buffer performance to cover the inevitable throughput drop during external load spike. Dynamic tuning (Algorithm~\ref{algo:dynamic_tuning_distributd}) for different SLAs are explained below.

\IncMargin{1em}
\begin{algorithm}[t]
\small
	\SetKw{in}{in}
	\SetKwData{Left}{left}\SetKwData{This}{this}
    \SetKwData{Up}{up}\SetKwData{Log}{Log}
    \SetKwData{ismedian}{$e_{s,median}$}\SetKwFunction{Median}{Median}
    \SetKwFunction{Union}{Union}
    \SetKwFunction{FindCompress}{FindCompress}
    \SetKwFunction{cluster}{cluster}
    \SetKwFunction{aggregateRequests}{aggregate\_requests}
    \SetKwFunction{networkStatus}{network\_view}
    \SetKwFunction{existingTransferStatus}{existing\_transfer\_status}
    \SetKwFunction{computeExternalLoad}{compute\_external\_load}
    \SetKwFunction{sendParamEnd}{send\_parameters}
    \SetKwFunction{estimateNewParam}{get\_params}
    \SetKwFunction{distributeParams}{distribute\_params}
    \SetKwFunction{distributeMaxMin}{distribute\_max\_min\_fair}
    \SetKwFunction{FindClosestSurface}{FindClosestSurface}
    \SetKwFunction{getParam}{get\_params}
    \SetKwFunction{GetOptimalParam}{GetOptimalParam}
    \SetKwFunction{DataTransfer}{DataTransfer}
    \SetKwFunction{AdaptiveSampling}{AdaptiveSampling}
    \SetKwFunction{append}{append}
    \SetKwFunction{remove}{remove}
    \SetKwFunction{QueryDB}{QueryDB}
    \SetKwFunction{Sort}{Sort}
    \SetKwFunction{GetSamples}{GetSamples}
	\SetKwInOut{Input}{input}\SetKwInOut{Output}{output}
	\Input{Transfer request, $req = \{src, dest, SLA\}$, \\
    	   Link information, $L[1:all]$, where,\\ $L_[i] = \{BW, RTT, T_{exist}\}$
    }
    
	\BlankLine

    \textbf{Periodically receive:} \\
    \networkStatus{} \\
    $ledger$ $\leftarrow$ \existingTransferStatus{} \\
    \BlankLine
    
    \textbf{if} request, $req$ is received \bf{:} $\quad$ $Queue.put(req)$
    
   	\While{Queue is not empty}{
    	$T_{ext}$ $\leftarrow$ \computeExternalLoad{$ledger$} \\
        \uIf{ req.SLA == 'Type-1'}{
        	$E_{exp}$, $\theta$ $\leftarrow$ \getParam{$Link$, $T_{ext}$, $T_{sla}$} 
        }
        \uElseIf{req.SLA == 'Type-2'}{
			$T_{exp}$, $\theta$ $\leftarrow$ \getParam{$Link$, $T_{ext}$, $E_{sla}$}
        }
        \Else{
        	$T_{exp}$, $\theta$ $\leftarrow$ \getParam{$Link$, $T_{ext}$, $\varphi_{sla}$}
        }
        \sendParamEnd{$\theta$} 
     }
     
\caption{Centralized Scheduling}
\label{algo:centralized_scheduling}
\end{algorithm}\DecMargin{1em}

\IncMargin{1em}
\begin{algorithm}[t]
\small
	\SetKw{in}{in}
    \SetKw{ORR}{or}
	\SetKwData{Left}{left}\SetKwData{This}{this}
    \SetKwData{Up}{up}\SetKwData{Log}{Log}
    \SetKwData{ismedian}{$e_{s,median}$}\SetKwFunction{Median}{Median}
    \SetKwFunction{Union}{Union}
    \SetKwFunction{centralizeScheduling}{centralize\_scheduling}
    \SetKwFunction{scaleDownParams}{scale\_down\_params}
    \SetKwFunction{FindCompress}{FindCompress}
    \SetKwFunction{cluster}{cluster}
    \SetKwFunction{microTune}{micro\_tune}
    \SetKwFunction{aggregateRequests}{aggregate\_requests}
    \SetKwFunction{estimateNewParam}{get\_params}
    \SetKwFunction{distributeParams}{distribute\_params}
    \SetKwFunction{distributeMaxMin}{distribute\_max\_min\_fair}
    \SetKwFunction{redistributeParams}{redistribute\_params}
    \SetKwFunction{FindClosestSurface}{FindClosestSurface}
    \SetKwFunction{getParam}{get\_params}
    \SetKwFunction{GetOptimalParam}{GetOptimalParam}
    \SetKwFunction{DataTransfer}{DataTransfer}
    \SetKwFunction{AdaptiveSampling}{AdaptiveSampling}
    \SetKwFunction{append}{append}
    \SetKwFunction{remove}{remove}
    \SetKwFunction{QueryDB}{QueryDB}
    \SetKwFunction{Sort}{Sort}
    \SetKwFunction{GetSamples}{GetSamples}
	\SetKwInOut{Input}{input}\SetKwInOut{Output}{output}
	\Input{ Periodic updates, $U[1:all] = \{EP_{s}, EP_{d}, T, E, Q_{rtt}, Status \}$
    }
    
	\BlankLine
    \If {\textrm{link capacity reduced due to failure/maintenance}}{
    	\scaleDownParams{... , $BW$=$new\_capacity$}
    }
    \For{$u_{i}$ \in $U[1:all]$}{
    	\If{ $u[status]$ == 'FINISHED' \ORR 'ABORTED' }{
        	\redistributeParams{}
        }
        \If{$u[status]$ == 'SLA violation'}{
        	\microTune{}
        }
        
    }

     
\caption{Centralized Micro-Tuning}
\label{algo:Micro-Tuning}
\end{algorithm}\DecMargin{1em}

\vspace{1mm}
\noindent {\textbf{Constraint 1 -- Throughput Guarantee:}} (Line 16-23)
It periodically checks the throughput, $T_{act}$ and checks whether the value is outside a tolerance bound $T_{sla} \pm \epsilon$. When it is below the bound, dynamic tuning initiates back-off control (Line 17-19), otherwise, it initiates \texttt{opportunistic\_increase()} module. 
In an uncongested link, it can provide continuous throughput guarantee, however, a congested link may force the transfer to reduce the parameters (ensure fairness) that directly impact the throughput. 

\texttt{opportunistic\_increase()} compensates this inevitable throughput drop with a new throughput goal $T_{goal}$ that is higher than $T_{sla}$ (Line 20-21). During uncongested time interval dynamic tuning module will ask offline analysis module to provide parameters that can achieve $T_{goal}$ (Line 22). 
We also introduce a buffer capacity of achievable throughput higher than $T_{sla}$. This buffer size is based on historical data analysis. Opportunistic increase function tries to fill this buffer, whenever it senses available throughput. This buffer pro-actively offsets any future throughput degradation. 

\begin{figure}[t]
\begin{centering}
\includegraphics[keepaspectratio=true,angle=0,width=87mm]{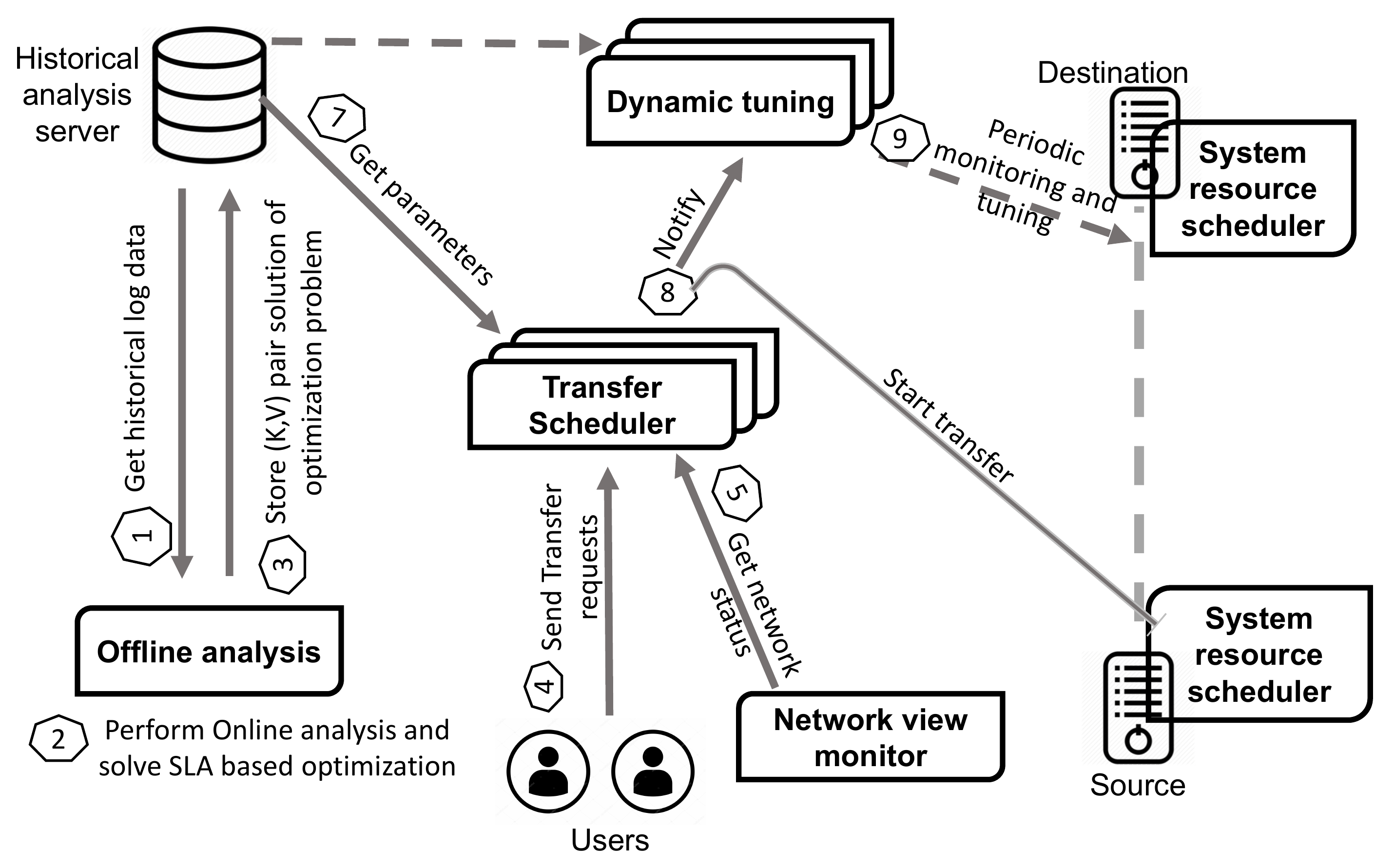}
	\end{centering}
 \caption{Centralized approach overview.} 
\label{fig:centralized_overview}
\end{figure} 

\vspace{1mm}
\noindent {\textbf{Constraint 2 -- Energy Constraint:}} (Line 6-15)
We have used a similar strategy for the energy constrained tuning as well. Here we are mostly interested in energy efficiency of the transfer, $\eta = Data_{total}/E_{total}$. This is the expected efficiency to maintain the energy constraint as mentioned in SLA. When the current efficiency goes below the expected efficiency, that means the transfer is using more energy to transfer unit amount of data than expected. It could be due to the congestion in the link where retransmission and waiting for acknowledgment can reduce $\eta$ significantly. We can easily check whether this is the case, just by measuring current throughput and queuing delay of the network and initiate the back-off control to reduce the parameters (Line 7-10). Another reason could be the unnecessary use of application-level parameters. Redundant concurrency levels could open more server processes which would lead to extra energy consumption. In such case, we ask offline analysis to provide new parameters. 
We also introduced opportunistic decrease of energy consumption whenever possible to offset the extra energy consumption happened due to suboptimal parameters or congestion. It can be done by switching processes to more energy constraint resource groups.

\vspace{1mm}
\noindent {\textbf{Constraint 3 -- Power Constraint:}} (Line 24-32)
This constraint asks to limit instantaneous power consumption of the transfer. However, in case of heavy congestion, it reduces the parameters. In a congestion-free network, it tries to increase the concurrency without violating the instantaneous power consumption limit.

\subsection{Centralized Approach}

The main advantage of centralized scheduling is the scheduler has a global view of the network status and overall transfer load. Therefore, the external load can be estimated more precisely. In distributed scheduling, there may be parameter over-shoot and under-shoot before all of them converge to an optimal level. However, this oscillation can be reduced in centralized approach, as the scheduler has global network view and all transfer periodically send status to the scheduler. An overview of the centralized approach is shown in Figure~\ref{fig:centralized_overview}. The centralized approach has three components: (1) offline analysis, (2) transfer scheduling, and (3) dynamic tuning. Transfer scheduler performs the offline historical analysis similar to distributed approach and pre-computes the optimization problems to use them during real-time transfers. 
 
After explaining away the external load, the available link bandwidth should not exceed the combined throughput guarantee of Type-1 SLA and combined predicted throughput of Type-2 SLA and Type-3 SLA. 
This module can work as an integration in  Traffic Engineering (TE) module of SDN. To make the solution more scalable, we can delegate controls in a hierarchical manner. 

\begin{table}[t]
\small
	\begin{centering}
    		\begin{tabular}{ |c|c|c|}
			\hline
			{\textbf{Specifications}} &  {\textbf{IBM IDCN} }  & {\textbf{XSEDE} }  \\
			\hline

			Bandwidth (GBps)	&  1 	 & 10	\\ \hline
			RTT (ms)	&   65	& 40 	\\	\hline
            Buffer size (MB)    &   8 & 32 \\ \hline
            File system &    & Lustre \\ \hline
           	Cores    &  2  & 16  \\  \hline
            Memory (GB)  & 2 & 32  \\ 
            \hline
            
		\end{tabular}
	    \caption{System and network specification of test sites}
        \label{tab:spec}
	\end{centering}
\end{table}

\subsubsection{Centralized Transfer Scheduler}

Initially, the scheduler (Algorithm~\ref{algo:centralized_scheduling}) clusters all the transfer requests based on source, destination, and their SLA requirements. Then it aggregates SLA requirements of each cluster. 
As we can see, each of the SLA groups in a single link is actually the external load for one another. The scheduler receives periodic updates from the participating transfers and has a more precise knowledge about the parameter distribution. It periodically updates external load estimation (Line 6).
Therefore, the centralized approach can ask offline analysis module for parameters with precise external load (Line 6-13), unlike distributed approach that starts with parameters for median external load and then converges. 

 \begin{figure*}[t]
    \begin{centering}
\begin{subfigure}[t]{0.32\textwidth}
        \includegraphics[keepaspectratio=true,width=50mm]{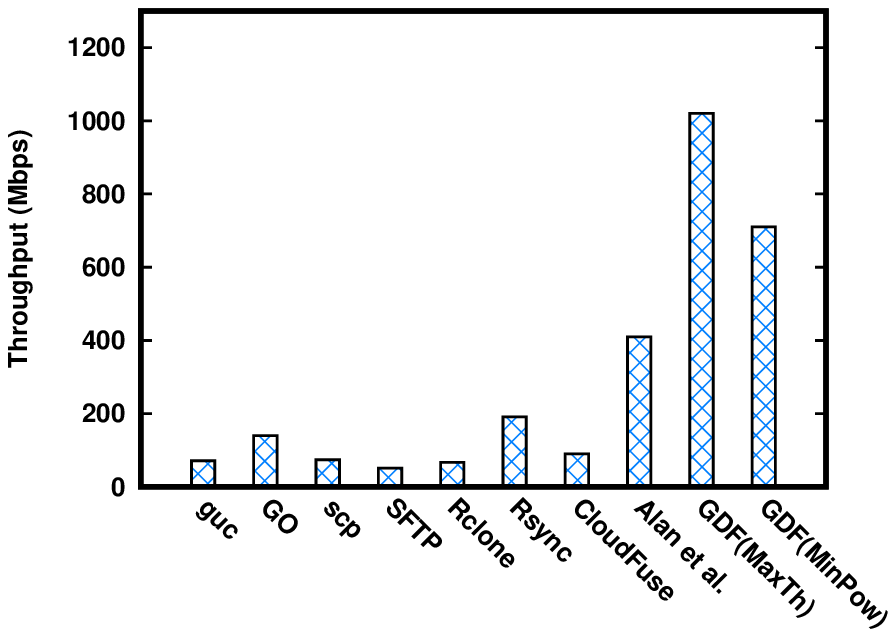}
        \caption{Achieved throughput (small files)}
    \end{subfigure}
    \begin{subfigure}[t]{0.32\textwidth}
        \includegraphics[keepaspectratio=true,width=50mm]{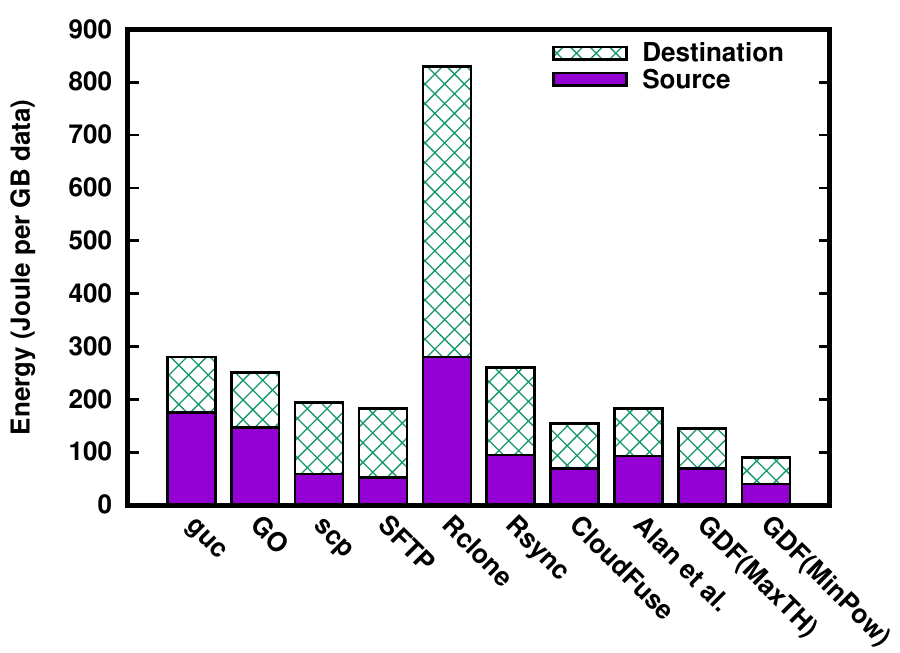}
        \caption{Energy consumption (small files)}
    \end{subfigure}
    \begin{subfigure}[t]{0.32\textwidth}
        \includegraphics[keepaspectratio=true,width=50mm]{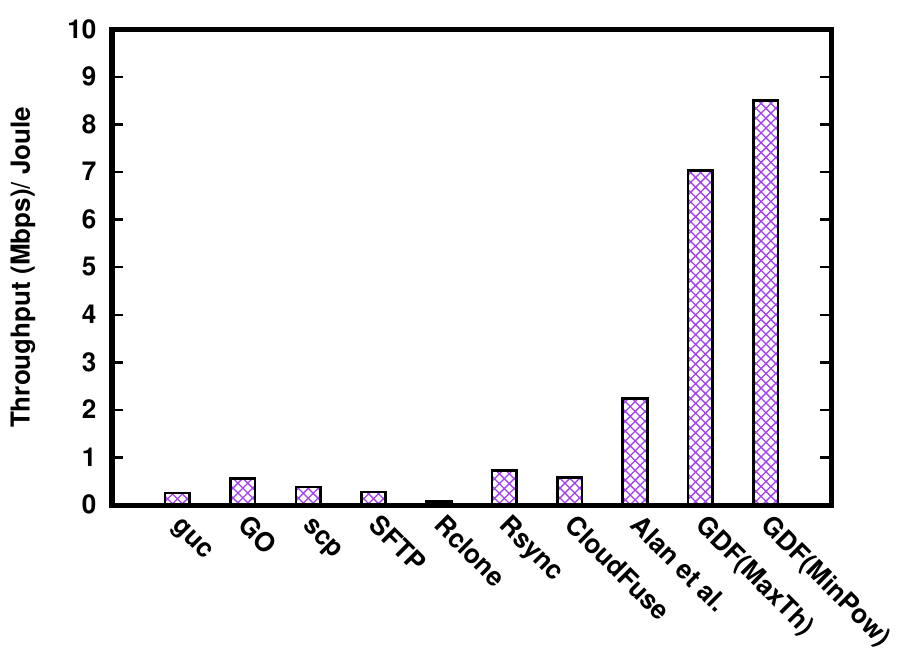}
        \caption{Throughput efficiency (small files)}
    \end{subfigure}
\begin{subfigure}[t]{0.32\textwidth}
        \includegraphics[keepaspectratio=true,width=50mm]{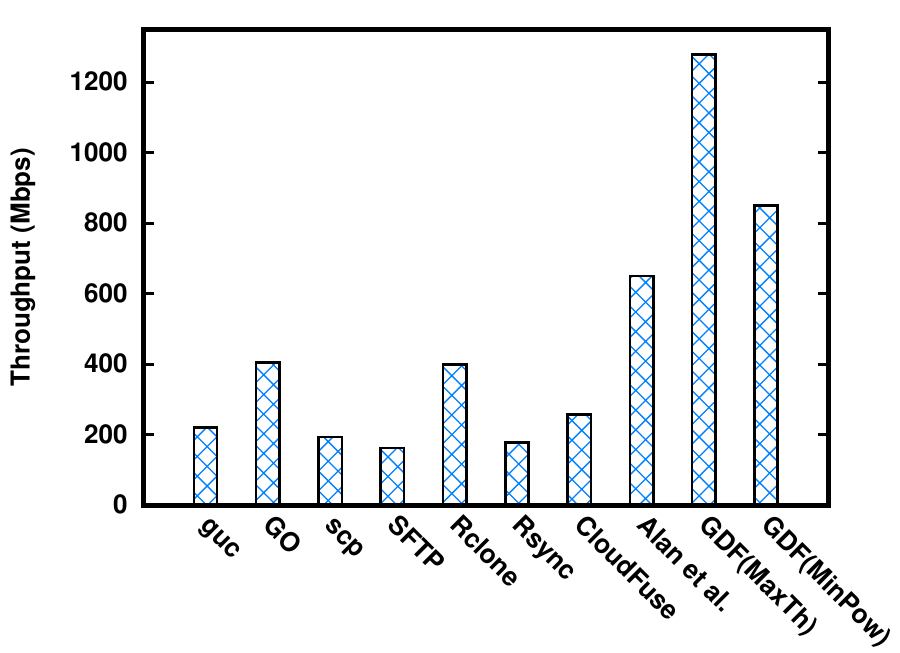}
        \caption{Achieved throughput (medium files)}
    \end{subfigure}
 \begin{subfigure}[t]{0.32\textwidth}
        \includegraphics[keepaspectratio=true,width=50mm]{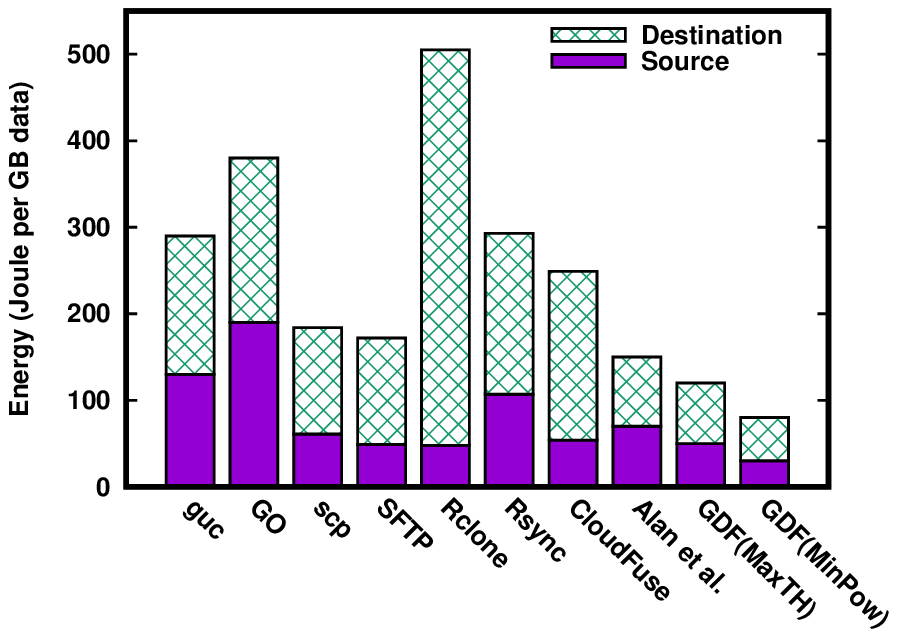}
        \caption{Energy consumption (medium files)}
    \end{subfigure}
    \begin{subfigure}[t]{0.32\textwidth}
        \includegraphics[keepaspectratio=true,width=50mm]{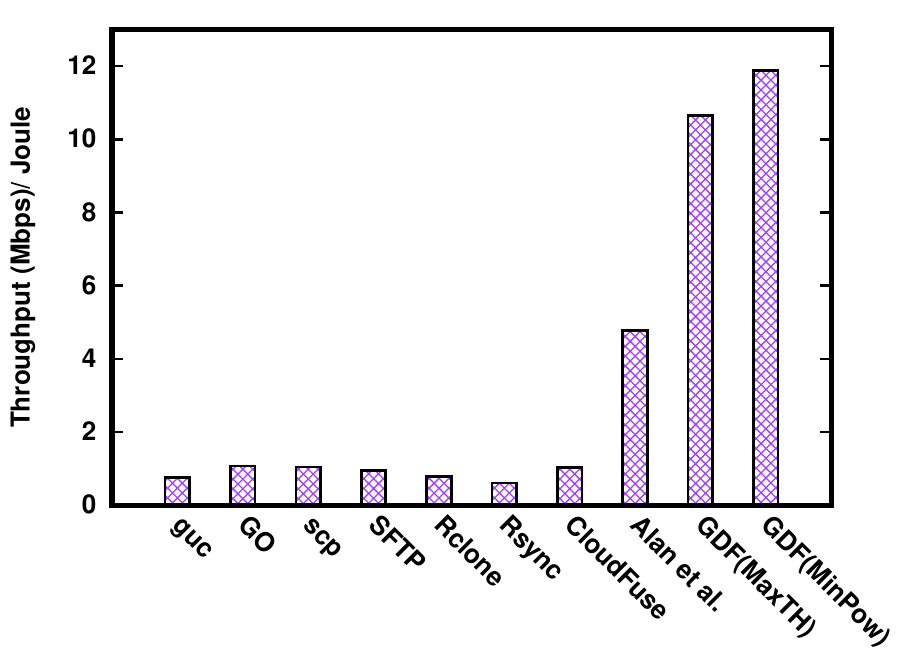}
        \caption{Throughput efficiency (medium files)}
    \end{subfigure}
\begin{subfigure}[t]{0.32\textwidth}
        \includegraphics[keepaspectratio=true,width=50mm]{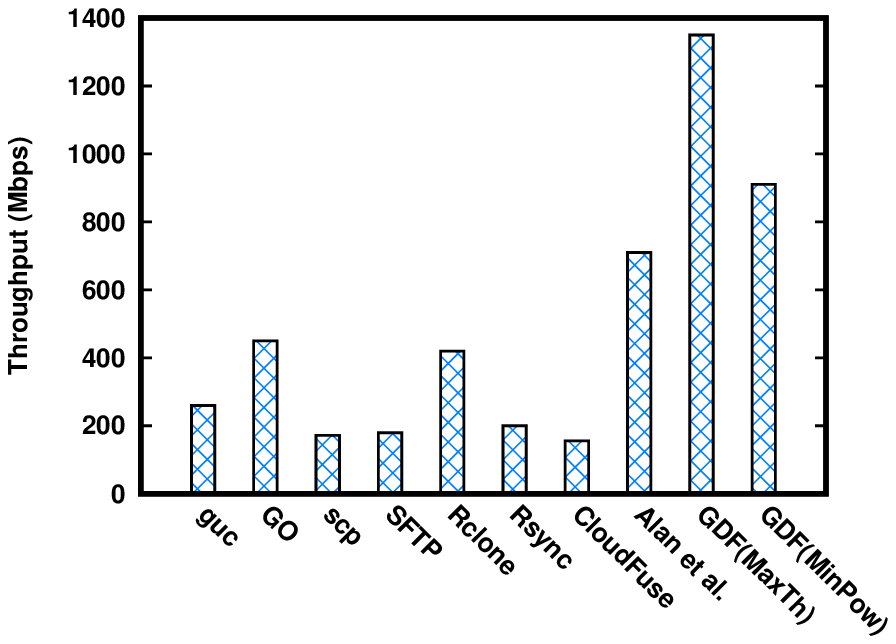}
        \caption{Achieved throughput (large files)}
    \end{subfigure}
    \begin{subfigure}[t]{0.32\textwidth}
        \includegraphics[keepaspectratio=true,width=50mm]{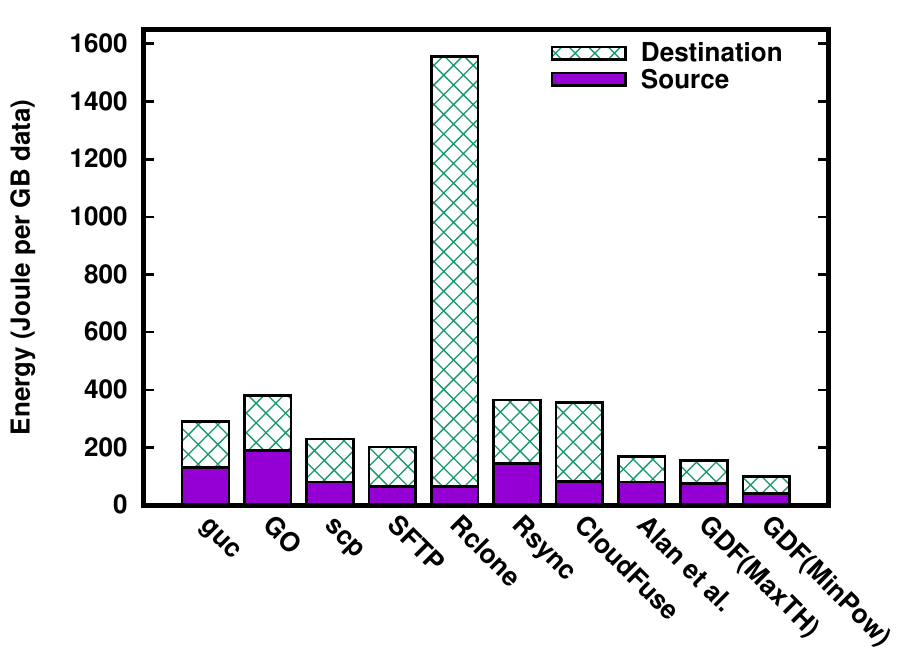}
        \caption{Energy consumption (large files)}
    \end{subfigure}
    \begin{subfigure}[t]{0.32\textwidth}
        \includegraphics[keepaspectratio=true,width=50mm]{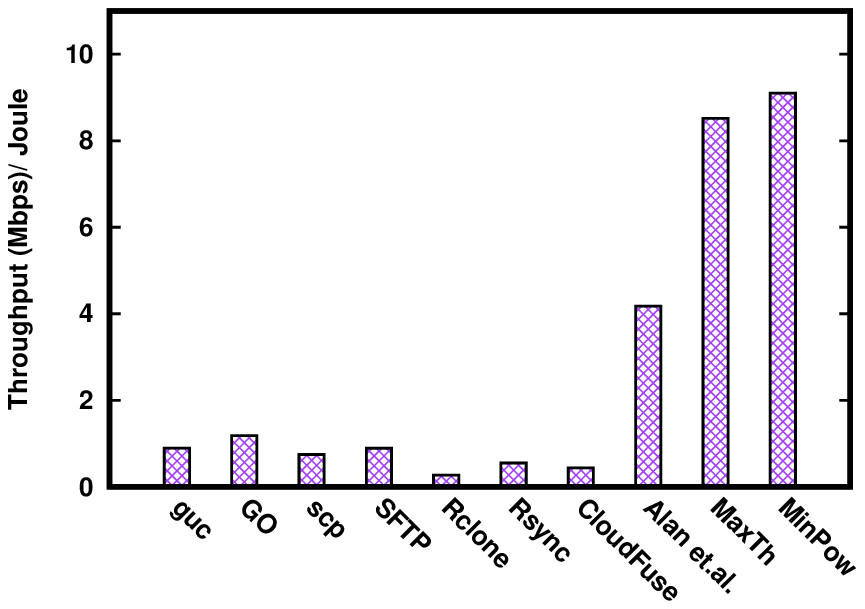}
        \caption{Throughput efficiency (large files)}
    \end{subfigure}
     \caption{Achievable throughput and corresponding energy consumption of different optimization objectives.}
     \label{fig:MFT_comparison}
     \end{centering}
 \end{figure*}
 
The centralized scheduler also performs micro tuning periodically when necessary (Algorithm~\ref{algo:Micro-Tuning}). 
In case of link capacity reduction due to failure or maintenance, scheduler scales down the parameters for all contending transfers.  
When a transfer finishes or is aborted, it notifies the scheduler so that it can redistribute the newly released parameters to existing transfers without violating energy and power constraints. In case of SLA violation, it redistributes the parameters among the contending transfer using more simpler \texttt{micro\_tune()} routine.

\section{Evaluation}
\label{sec:experiments}
We performed experiments on a wide-area network link between IBM datacenters located in Washington, D.C. and San Jose, CA. We also used XSEDE, a production level high-speed computing infrastructure for large-scale scientific computations. An overview of systems and network information is provided in Table~\ref{tab:spec}. 

We compared our model with many existing data transfer solutions. However, there has been done very little work on energy efficient data transfer optimization. We evaluate our model against the model proposed by Alan et. al.~\cite{Alan2015}, \texttt{globus-url-copy} (guc)~\cite{foster1997globus}, Globus Online (GO)~\cite{chard2017globus}, \texttt{scp}, \texttt{SFTP}, Rsync, Rclone~\cite{Rclone@2017}, and CloudFuse~\cite{CloudFuze@2016}. 

Alan et. al. provide a High Throughput Energy-Efficient Transfer Algorithm (HTEE) that uses heuristics based approach to balancing the achieved throughput and energy consumption. It starts with one channel and periodically increases it by 2 until it reaches to a user-defined limit. Then it computes ($T_{act}/E_{act}$) ratio for each level and picks the best one. This periodic additive increase is slow and keeps the transfer sub-optimal until it searches the whole parameter space which is still $O(n)$ when the user-defined value is $n$. 
\texttt{Globus-url-copy} and Globus Online are GridFTP based data transfer tools to achieve high performance during transfer. However, they are not energy optimized tools. Scp and SFTP are widely used secure file transfer tools. Rsync and Rclone are high-performance data synchronization applications. CloudFuse provides cloud-based Managed File Transfer (MFT) service and offers migration, sync and other file management capabilities to the end users. 

\subsection{Comparison with Other Solutions}
Figure~\ref{fig:MFT_comparison} shows an elaborate experimentation and performance analysis of different state-of-the-art solutions and our proposed approach. In literature, very little work is done to optimize both throughput and energy of a data transfer with SLA specifications. Most of the models do not support SLA. Therefore, to make comparison fair we set the SLA of our model in two extreme cases - (1) Maximum achievable throughput (MaxTh) and (2) Minimum possible energy consumption (MinPow). To test the efficiency of different types of file transfers we tested all data transfer solutions for small (1 - 5MB), medium (100 - 500MB), and large (1 - 4GB) files. 

\subsubsection{Performance of Medium File Transfers}
Figure~\ref{fig:MFT_comparison} (d-f) contain performance comparison for medium files. We compared achieved throughput, energy consumption, and the throughput efficiency, ($T_{act}/E_{act}$). To accelerate the medium file transfer, a moderate choice of concurrency $cc$ and parallelism $p$ is helpful in an uncongested link. Multiple files with multiple segments can be transferred simultaneously. However, an initial best-known parameter choice along with dynamic tuning and effective end-system resource scheduling can increase energy efficiency as well. 

As we can see off-the-shelf tools like scp and SFTP perform poorly due to single data channel allocation and also the control channel inefficiency in long RTT WAN. Their energy consumption is also high because low throughput transfer needs more time to finish and even though resource utilization is low, longer time of execution increases the static power component (power consumption when the resource is idle and waiting). Similarly, \texttt{globus-url-copy (guc)}, a GridFTP based tool, also performs poorly with base-line parameter settings ($cc=1$ \& $p=1$). 

GridFTP is designed for multi-threaded transfers and more resource intensive than \texttt{scp} and \texttt{SFTP} when used with a single channel. So, it suffers from low throughput while consuming more energy. On the other hand, Globus Online is a statically tuned cloud service that uses GridFTP protocol. Due to the use of multiple streams ($cc\times p > 1$), we observe that it can reach up to $2\times$ performance improvement compare to \texttt{scp}, \texttt{SFTP,} and \texttt{guc}. However, it consumes $2\times$ more energy. The reason is that statically assigned parameters may not be optimal for all external traffic levels. There is no way to limit the resource utilization as well.  

\texttt{Rclone} and \texttt{Rsync} are file synchronization tools. By default, \texttt{Rclone} uses 4 parallel data connections to transfer a single file. We observe that it can achieve similar performance as \texttt{GO}, however, there is no way to do any dynamic adjustment to parallel connections. We see an unusually high energy consumption at the destination. It may be due to the extra work it has to do for sync operation for four parallel connections. On the other hand, \texttt{Rsync} performance is lower than \texttt{Rclone} as it does not use any parallel connections for file transfer, therefore, under-utilizes the available bandwidth. However, it consumes less energy due to its single connection syncing. 

CloudFuse achieves slightly better performance than \texttt{guc}. It consumes slightly less energy compared to  \texttt{Rsync} and \texttt{guc}. As we see in the Figure, Alan et. al. model performs much better than solutions discussed above due to the fact that it performs an on-line parameter search and after deciding on the best parameter, it transfers rest of the data efficiently. However, there is no real-time control on parameters, therefore, when external traffic changes those parameters may become sub-optimal. And additive parameter search may take a toll on the achieved throughput. However, due to the search for energy-efficient parameters, it can manage to keep energy consumption less than other approaches mentioned above.

Both of GreenDataFlow algorithms (MaxTh and MinPow) outperform all the listed solutions. MaxTh provides $6\times$ throughput performance improvement over the baseline performance of \texttt{globus-url-copy} and almost $2\times$ improvement over the closest competitor Alan et al. model due to the historical analysis and real-time tuning of the parameters. As it achieves high throughput, the execution time reduces as well which reduces the static power consumption along with constrained resource scheduling in end-systems. MinPow is aimed to decrease the total energy consumption. It consumes $8\times$ less energy than the energy-hungry \texttt{rclone}, $3\times$ less energy than base-line \texttt{guc} and almost 36\% less energy than the closest competitor Alan et. al.  
  
\subsubsection{Performance of Small and Large File Transfers}
Disk-to-disk small file transfers are very challenging, as we have to perform a lot of disk read/write operations. Without a proper pipelining level, control channel idleness can introduce delay among subsequent file transfers. As we can see in Figure~\ref{fig:MFT_comparison} (a-c), overall achieved throughput for all approaches are lower than the achieved throughput of medium (Figure~\ref{fig:MFT_comparison}(d)) and large files (Figure~\ref{fig:MFT_comparison}(g)). However, the achievable throughput difference is quite similar to medium file transfers except the cases where \texttt{Rclone} performs worst than \texttt{Rsync} and \texttt{CloudFuse}. Alan et al. model achieves low throughput compared to medium files, because small file transfers suffer badly for sub-optimal parameter choices. Therefore, additive parameter search takes more toll on small file transfers compared to the medium files. Our MaxTh model reaches $2.5\times$ performance increase compared to the closest competitor Alan et al. model, and our energy optimized MinPow consumes $2\times$ less energy compared to it.

 \begin{figure*}[t]
    \begin{centering}
    \begin{subfigure}[t]{0.30\textwidth}
        \includegraphics[keepaspectratio=true,width=41mm]{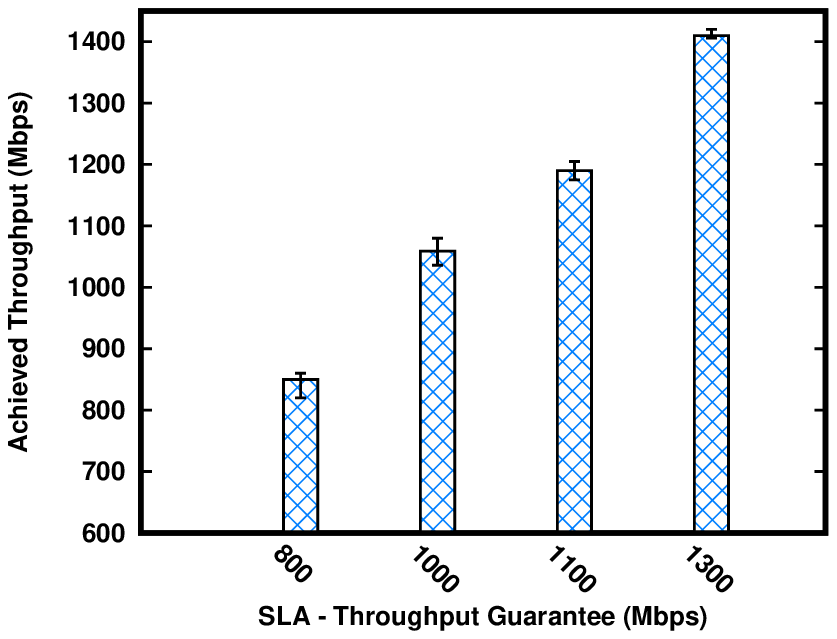}
        \caption{Achieved Throughput (SLA Type-T)}
    \end{subfigure}
    \begin{subfigure}[t]{0.30\textwidth}
        \includegraphics[keepaspectratio=true,width=41mm]{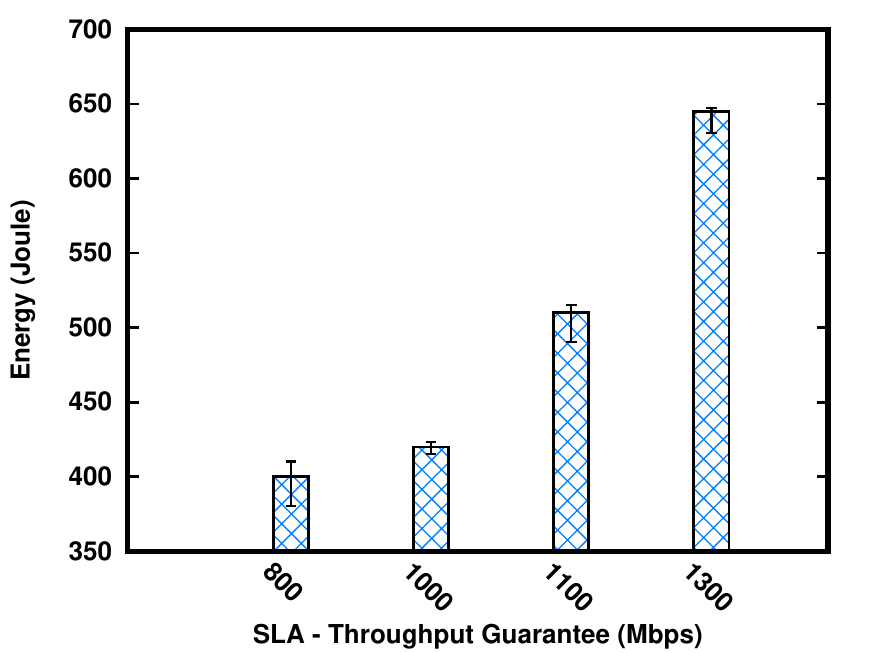}
        \caption{Energy Consumption (SLA Type-T)}
    \end{subfigure}
    \begin{subfigure}[t]{0.30\textwidth}
        \includegraphics[keepaspectratio=true,width=41mm]{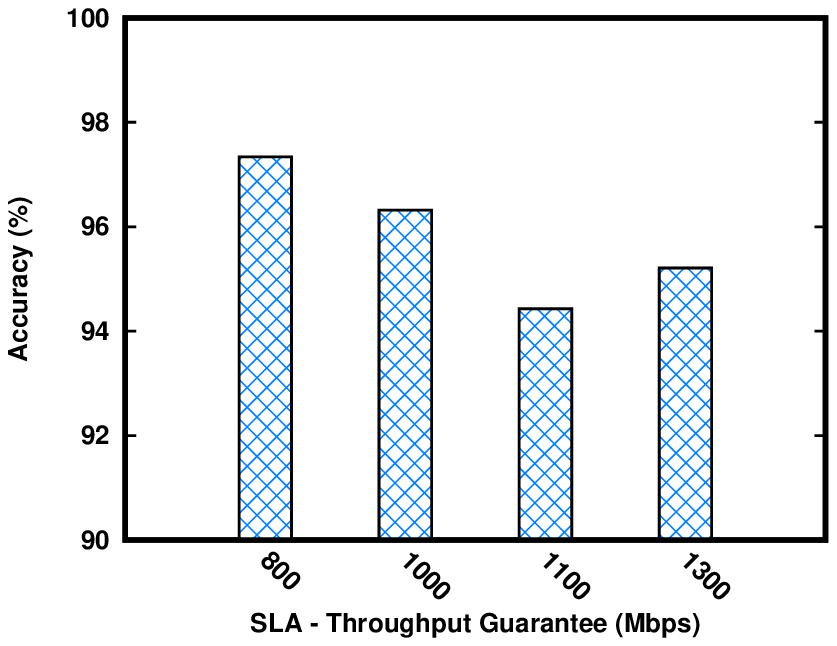}
        \caption{SLA Accuracy (SLA Type-T)}
    \end{subfigure}
    \begin{subfigure}[t]{0.30\textwidth}
        \includegraphics[keepaspectratio=true,width=41mm]{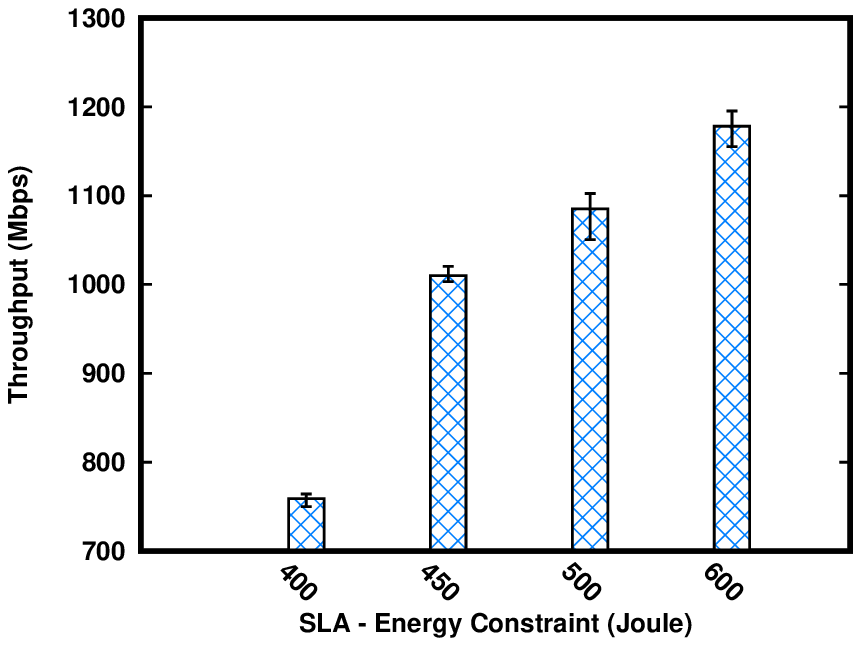}
        \caption{Achieved Throughput (SLA Type-E)}
    \end{subfigure}
    \begin{subfigure}[t]{0.30\textwidth}
        \includegraphics[keepaspectratio=true,width=41mm]{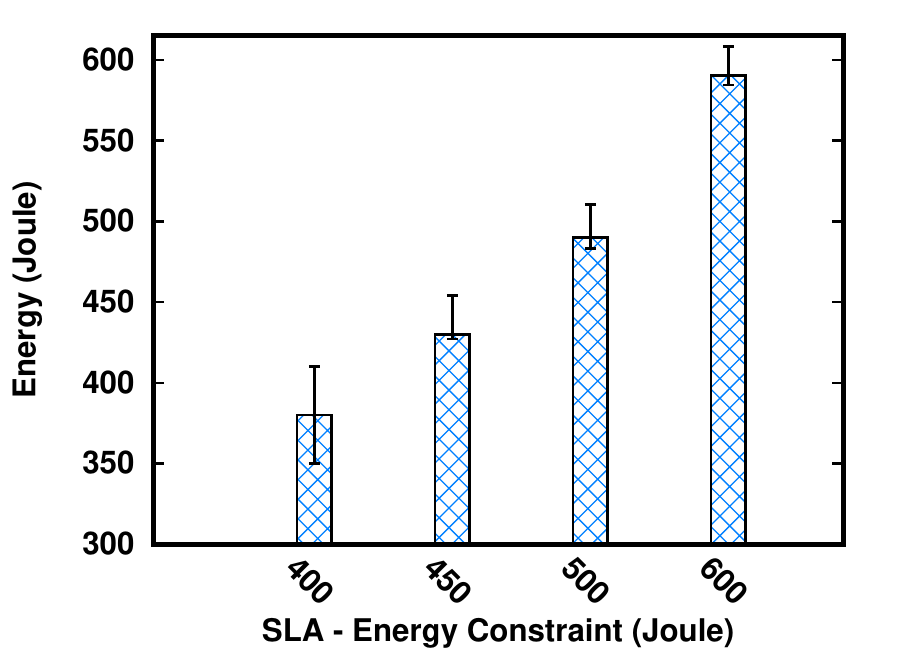}
        \caption{Energy Consumption (SLA Type-E)}
    \end{subfigure}
\begin{subfigure}[t]{0.30\textwidth}
        \includegraphics[keepaspectratio=true,width=41mm]{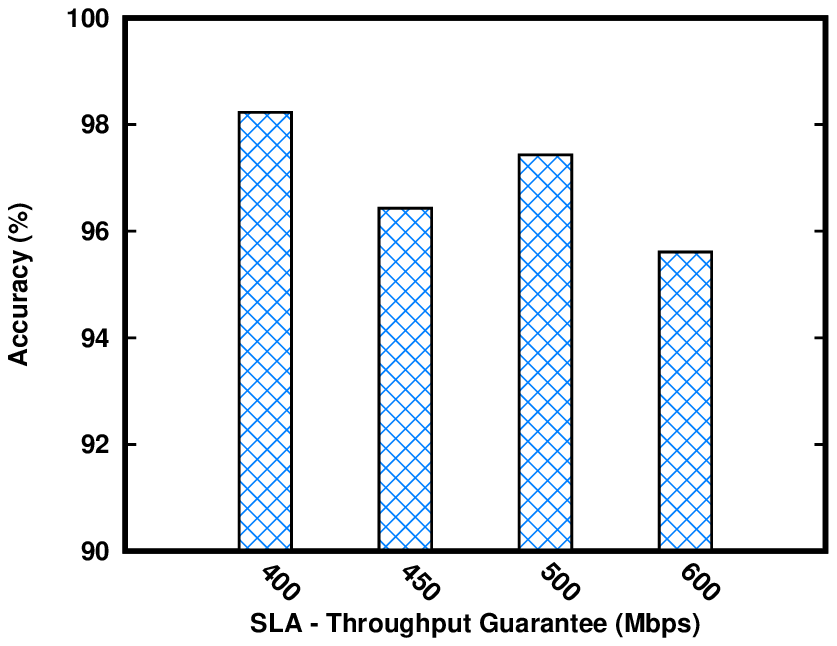}
        \caption{SLA Accuracy (SLA Type-E)}
    \end{subfigure}
    \begin{subfigure}[t]{0.30\textwidth}
        \includegraphics[keepaspectratio=true,width=41mm]{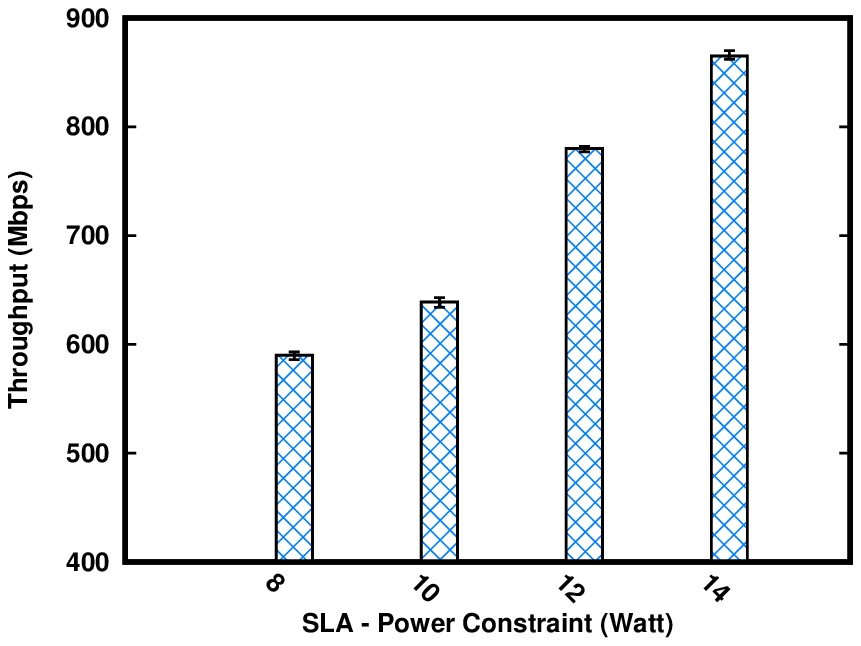}
        \caption{Achieved Throughput (SLA Type-P)}
    \end{subfigure}
    \begin{subfigure}[t]{0.30\textwidth}
        \includegraphics[keepaspectratio=true,width=41mm]{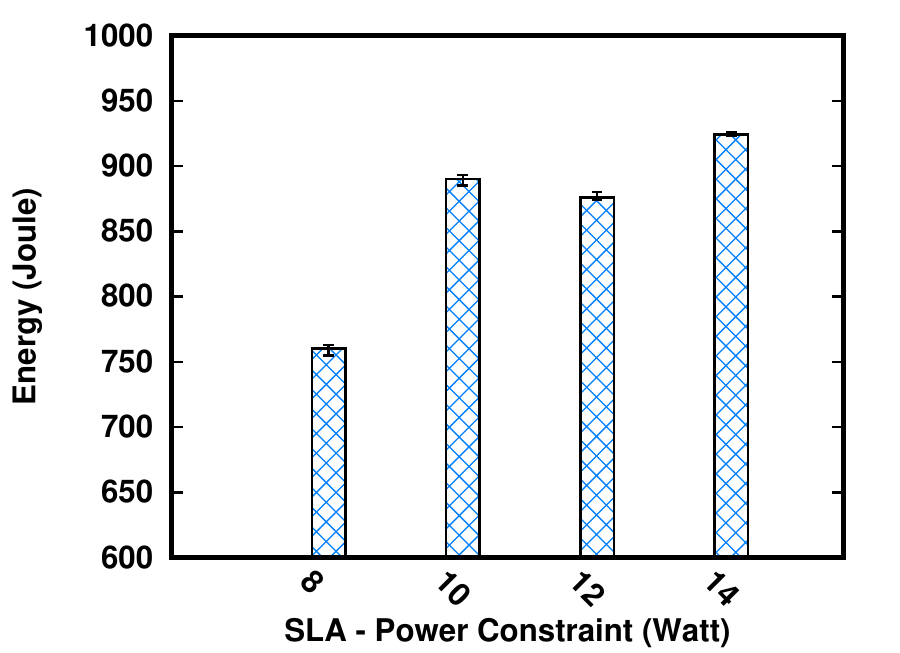}
        \caption{Energy Consumption (SLA Type-P}
    \end{subfigure}
    \begin{subfigure}[t]{0.30\textwidth}
        \includegraphics[keepaspectratio=true,width=41mm]{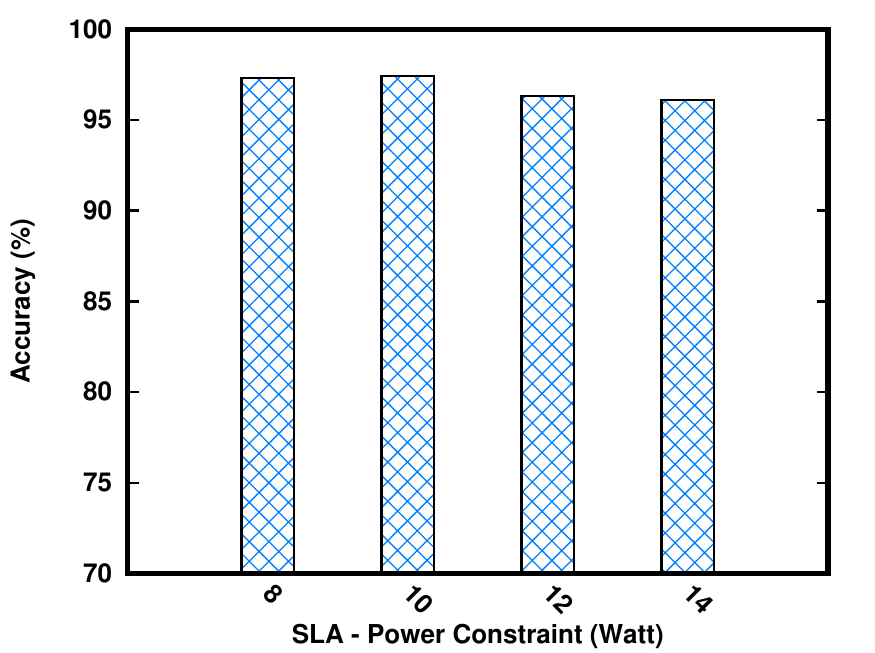}
        \caption{SLA Accuracy (SLA Type-P)}
    \end{subfigure}
     \caption{Achieved throughput, energy consumption and SLA commitment accuracy for different SLA types. (a-c) SLA with throughput guarantee, (d-f) SLA with energy constraint, and (g-i) SLA with power constraint.}
     \label{fig:SLA constraint}   
     \end{centering}
 \end{figure*}

Parallelism is the most important parameter for large file transfers as we want to parallelize multiple segments of a large file. Concurrency can add extra boost on performance, however, a very high value can over-burden the network. Figure~\ref{fig:MFT_comparison} (g-i) shows the performance of large file transfers. It can be seen that the overall throughput performance is better than medium file transfers. However, the performance pattern for \texttt{guc, GO, SCP, SFTP, rsync, CloudFuse} is roughly similar to medium files because of the use of fixed parameter settings. Energy consumption performance is also similar to medium files except for a huge spike in \texttt{rclone} destination, as the file size grows the energy consumption increases rapidly. Alan et al. model also performs better compare to its performance on small and medium files. Our model outperforms Alan et al. model and achieves $2\times$ performance boost in throughput. On the other hand, our MinPow model, consumes 70\% less energy compare to Alan et al. model.  

\subsection{SLA-based Performance Analysis}
We have discretized the SLA levels of throughput, energy and instant power and eliminated the infeasible regions. Figure~\ref{fig:SLA constraint} shows the performance for different SLA levels. It can be seen that SLA violations  are rare unless there exist over-subscription of throughput or severe capacity reduction for a long period of time. Most of the cases in Type-T SLA (Figure~\ref{fig:SLA constraint} (a-c)), our model can achieve performance over $T_{sla}$, due to the \texttt{opportunistic\_increase} strategy. It also keeps the resource utilization manageable by putting dynamic restriction on usage. SLA violation error is ranged from 3\% to 6\%. For energy constrained (Type-E) SLA, we observed the SLA violation occurs due to heavy congestion which forces retransmission and initiates slow start phase. Moreover, excessive concurrent processes can consume extra power while congesting the network. As our model cautiously monitors and budgets the required future energy usage, it can achieve high accuracy in SLA commitment. Instant power consumption constraint forces the transfer to be assigned in a restricted resource group and never changes it to guarantee this constraint, however, it may fix parameters in real-time to achieve the expected throughput. This constraint produces some interesting results, as we can see, it consumes more total energy to achieve a throughput similar to Type-T and Type-E SLAs. Due to the strict resource constraint, transfer may take long time to finish that leads to more static power consumption.   

\subsection{Protocol Fairness Analysis}

\begin{figure}[t]
\begin{centering}
\includegraphics[keepaspectratio=true,angle=0,width=70mm]{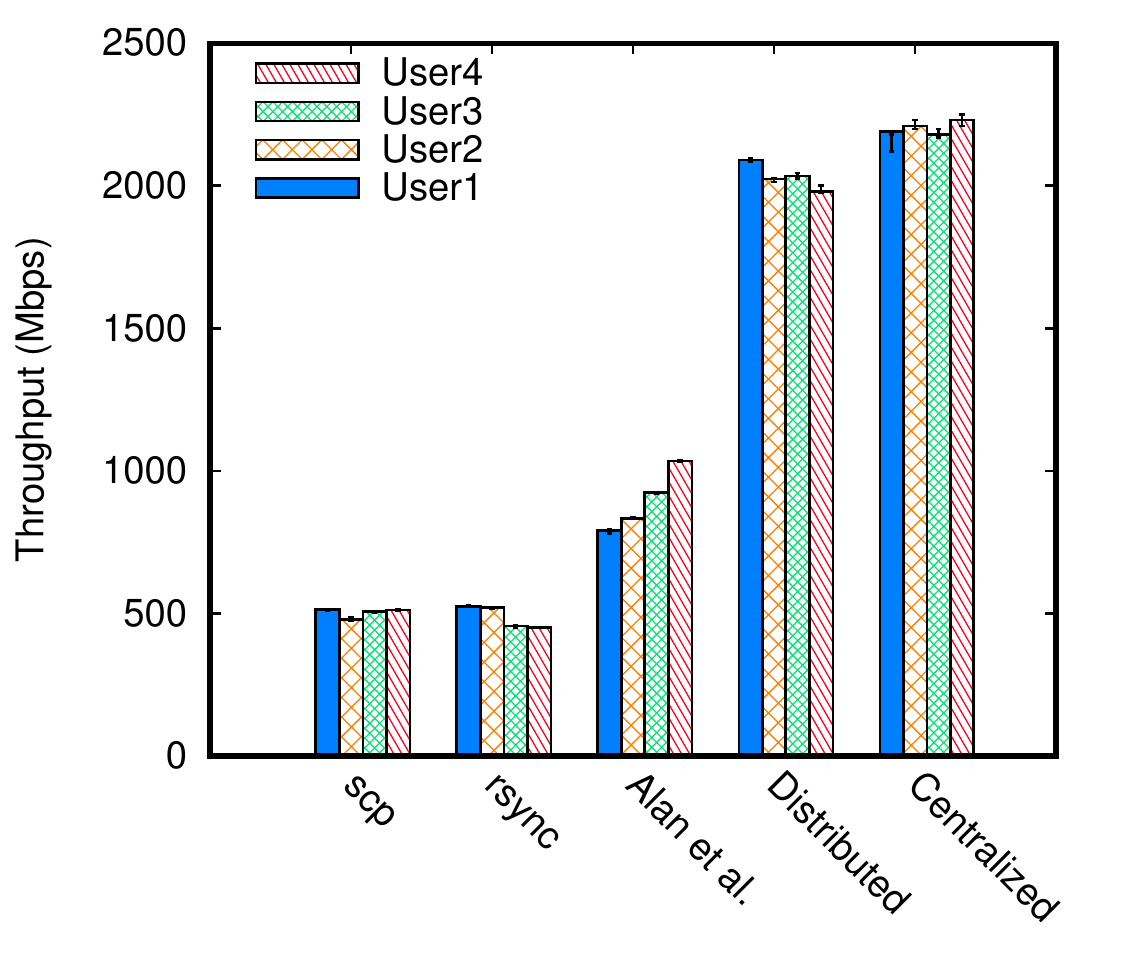}
	\end{centering}
\caption{Fairness analysis of different approaches.} 
\label{fig:fairness_analysis}
\end{figure}

Both of our distributed and centralized approaches are designed to maintain fairness among the contending transfers while maximizing the overall WAN utilization. We have tested our model in 10 Gbps XSEDE WAN with four contending transfers. Figure~\ref{fig:fairness_analysis} shows the performance of different transfer approaches. We can see that \texttt{scp} can achieve throughput around 500 Mbps and all contending users can get an equal share, however, the network utilization is very low. \text{Rsync} also gets a similar performance with a good fairness among the users. This is due to the single channel that is not enough to fill up the 10 Gbps WAN link. Both of them achieve 20\% of the network utilization. Alan et al. model achieves higher throughput and the network utilization is almost 40\%, however, we can see the performance is not fairly distributed among the users. If all the transfer start at the same time, then Alan et al. model can get a fair share. If external traffic changes during the search process, then it can choose parameters unfairly. Both the distributed and centralized models provide superior utilization of the network. The distributed approach can achieve almost 82\% network utilization where centralized approach can reach up to 90\% utilization. In the distributed approach, the users need to sense the network periodically and waste some throughput while converging. However, centralized approach converges faster as it knows about other contending transfers and estimates external load more precisely. Due to the proper back-off control, it can become less aggressive towards the other contending transfers.

\section{Related Work}
\label{sec:Related Work}
The work on network throughput optimization focuses on tuning transfer parameters such as parallelism, pipelining, concurrency and buffer size.
The first attempts to improve the data transfer throughput at the application layer were made through buffer size tuning. Various dynamic and static methods were proposed to optimize the buffer size~\cite{jain2003tcp,prasad2003socket,semke1998automatic}. However, Lu et al.~\cite{lu2005modeling} showed that parallel streams can achieve a better throughput than buffer size tuning and then several others~\cite{Altman06paralleltcp,R_Hacker05,Yildirim:2009:BTB:1552280.1552283,JGrid_2012} proposed throughput optimization solutions by means of tuning parallel streams. Another transfer parameter used for throughput optimization was pipelining, which helped in improving the performance of transferring large number of small files~\cite{TCP_Pipeline,farkas2002,R_Bres07, NDM_2012}. Liu et al.~\cite{R_Liu10} optimized network throughput by concurrently opening multiple transfer sessions and transferring multiple files concurrently. They proposed increasing the number of concurrent data transfer  channels until the network performance degrades. Globus Online~\cite{globusonline} offers fire-and-forget file transfers through thin clients over the Internet. It partitions files based on file size and transfer each partition using partition-specific protocol parameters. However, the protocol tuning Globus Online performs is non-adaptive; it does not change depending on network conditions and transfer performance.

The work on power-aware networking focuses on saving energy at the networking devices. Gupta et al.~\cite{Gupta_2003} were among the earliest researchers to advocate conserving energy in networks. They suggested different techniques such as putting idle sub-components (i.e. line cards, etc.) to sleep~\cite{Gupta_2007}, which were later extended by other researchers. S. Nedevshi et al.~\cite{Nedevschi_2008} proposed adapting the rate at which switches forward packets depending on the traffic load. IEEE Energy Efficient Ethernet task force proposed the 802.3az standards ~\cite{IEEE_802} for making ethernet cards more energy efficient. They defined a new power state called low power idle (LPI) that puts the ethernet card to low power mode when there is no network traffic. Other related research in power-aware networking has focused on architectures with programmable switches~\cite{Greenberg_2008} and switching layers that can incorporate different policies~\cite{Joseph_2008}. Barford et al. proposed power-aware network protocols for energy-efficiency in network design and routing~\cite{Barford_2008}. 
Bertozzi et al.~\cite{bertozzi2003transport} investigated the energy trade-off in networking as a function of the TCP receive buffer size and show that the TCP buffering mechanisms can be exploited to significantly increase energy efficiency of the transport layer with minimum performance overheads.

Several highly-accurate scheduling algorithms~\cite{Thesis_2005, SciProg_2007, ICEIS_2007} and predictive models ~\cite{ICSIS_2009,R_Yin11, R_Yildirim11, DISCS12, Cluster_2015} were developed which require as few as three sampling points to provide very accurate predictions for the parallel stream number giving the highest transfer throughput for the wired networks. Yildirim et al. analyzed the combined effect of parallelism and concurrency on data transfer throughput~\cite{TCC2015}.
Alan et al. analyzed the effects of parallelism and concurrency on end-to-end data transfer throughput versus total energy consumption in wide-area wired networks using precalculated values for these parameters and proposed a heuristic approach to improve them~\cite{Alan2015, Kosar_jrnl14}.

\section{Conclusion}
\label{sec:conclusion}
In this paper, we introduced a novel set of data transfer algorithms (collectively called GrenDataFlow) based on historical analysis and real-time tuning, which can achieve high data transfer throughput while keeping the energy consumption during the transfers at the minimal levels. GreenDataFlow supports service level agreements (SLAs) which give the service providers and the consumers the ability to fine tune their goals in this optimization process. Our experimental results show that GreenDataFlow outperforms existing solutions in this area both in terms of energy saving and the achieved end-to-end performance. Considering the massive energy footprint of global data movement, our presented GreenDataFlow techniques have a great potential to decrease this footprint and contribute to the efforts of achieving greener Internet.

\section*{Acknowledgements}
This project is in part sponsored by the National Science Foundation (NSF) under award numbers OAC-1724898 and OAC-1842054, and by IBM under award number OCR-W1771224.

\bibliographystyle{abbrv}
\bibliography{main,didc,misc,rela_work,sc2017} 

\end{document}